\input harvmac.tex
\input epsf.tex
\def\figin{\epsfcheck\figin}\def\figins{\epsfcheck\figins}
\def\epsfcheck{\ifx\epsfbox\UnDeFiNeD
\message{(NO epsf.tex, FIGURES WILL BE IGNORED)}
\gdef\figin##1{\vskip2in}\gdef\figins##1{\hskip.5in}
\else\message{(FIGURES WILL BE INCLUDED)}%
\gdef\figin##1{##1}\gdef\figins##1{##1}\fi}
\def\DefWarn#1{}
\def\figinsert{\goodbreak\midinsert}
\def\ifig#1#2#3{\DefWarn#1\xdef#1{fig.~\the\figno}
\writedef{#1\leftbracket fig.\noexpand~\the\figno}%
\figinsert\figin{\centerline{#3}}\medskip\centerline{\vbox{\baselineskip12pt
\advance\hsize by -1truein\noindent\footnotefont{\bf Fig.~\the\figno:}
#2}}
\bigskip\endinsert\global\advance\figno by1}

\newdimen\tableauside\tableauside=1.0ex
\newdimen\tableaurule\tableaurule=0.4pt
\newdimen\tableaustep
\def\phantomhrule#1{\hbox{\vbox to0pt{\hrule height\tableaurule
width#1\vss}}}
\def\phantomvrule#1{\vbox{\hbox to0pt{\vrule width\tableaurule
height#1\hss}}}
\def\sqr{\vbox{%
  \phantomhrule\tableaustep

\hbox{\phantomvrule\tableaustep\kern\tableaustep\phantomvrule\tableaustep}%
  \hbox{\vbox{\phantomhrule\tableauside}\kern-\tableaurule}}}
\def\squares#1{\hbox{\count0=#1\noindent\loop\sqr
  \advance\count0 by-1 \ifnum\count0>0\repeat}}
\def\tableau#1{\vcenter{\offinterlineskip
  \tableaustep=\tableauside\advance\tableaustep by-\tableaurule
  \kern\normallineskip\hbox
    {\kern\normallineskip\vbox
      {\gettableau#1 0 }%
     \kern\normallineskip\kern\tableaurule}%
  \kern\normallineskip\kern\tableaurule}}
\def\gettableau#1 {\ifnum#1=0\let\next=\null\else
  \squares{#1}\let\next=\gettableau\fi\next}

\tableauside=1.0ex
\tableaurule=0.4pt

\def\inbar{\,\vrule height1.5ex width.4pt depth0pt}
\def\IC{\relax\hbox{$\inbar\kern-.3em{\rm C}$}}
\def\IR{\relax{\rm I\kern-.18em R}}
\def\IZ{\relax{\rm I\kern-.18em Z}}

\lref\cs{E. Witten, ``Quantum field theory and the Jones polynomial," 
Commun. Math. Phys. {\bf 121} (1989) 351.}
\lref\gvtwo{R. Gopakumar and C. Vafa, ``On the gauge theory/geometry 
correspondence," hep-th/9811131.}
\lref\ov{H. Ooguri and C. Vafa, ``Knot invariants and topological 
strings," hep-th/9912123, Nucl. Phys. {\bf B 577} (2000) 419.} 
\lref\cmr{S. Cordes, G. Moore and S. Ramgoolam, ``Large $N$ 
2D Yang-Mills theory and topological string theory,'' 
hep-th/9402107, Commun. Math. Phys. {\bf 185} (1997) 543. ``Lectures on  
two-dimensional Yang-Mills theory, equivariant cohomology, and 
topological field theory," hep-th/9412210, Nucl. Phys. Proc. Suppl. {\bf
41} 
(1995) 184.}
\lref\gv{R. Gopakumar and C. Vafa, ``M-theory and topological 
strings, I," hep-th/9809187. }
\lref\fp{C. Faber and R. Pandharipande, ``Hodge integrals and 
Gromov-Witten theory,'' math.AG/9810173, Invent. Math. {\bf 139} (2000)
173.}
\lref\gvm{R. Gopakumar and C. Vafa, ``M-theory and topological 
strings, II," hep-th/9812127. }
\lref\kkv{S. Katz, A. Klemm and C. Vafa, ``M-theory, topological strings,
and 
spinning black-holes," hep-th/9910181.}
\lref\vare{C. Vafa, ``Superstrings and topological strings at large $N$,'' 
hep-th/0008142.}
\lref\llr{J.M.F. Labastida, P.M. Llatas and A.V. Ramallo, ``Knot operators 
in Chern-Simons theory," Nucl. Phys. {\bf B 348} (1991) 651.}
\lref\lmtwo{J.M.F. Labastida and M. Mari\~no, ``The HOMFLY polynomial for 
torus links from Chern-Simons gauge theory," hep-th/9402093, Int. J. Mod.
 Phys. {\bf A 10} (1995) 1045.} 
\lref\ilr{J.M. Isidro, J.M.F. Labastida, and A.V. Ramallo, 
``Polynomials for torus links from Chern-Simons gauge theories," 
hep-th/9210124, Nucl. Phys. {\bf B 398} (1993) 187.}
\lref\fh{W. Fulton and J. Harris, {\it Representation theory. A first
course},
Springer-Verlag, 1991.}
\lref\jonesann{V.F.R. Jones, ``Hecke algebras representations of braid 
groups and link polynomials," Ann. of Math. {\bf 126} (1987) 335. }
\lref\guada{E. Guadagnini, ``The universal link polynomial," 
Int. J. Mod. Phys. {\bf A 7} (1992) 877;  {\it The link invariants of the 
Chern-Simons field theory,} Walter de Gruyter, 1993.}
\lref\lick{W.B.R. Lickorish, {\it An introduction to knot theory}, 
Springer-Verlag, 1998.}
\lref\awrep{M. Wadati, T. Deguchi and Y. Akutsu, ``Exactly solvable models 
and knot theory," Phys. Rep. {\bf 180} (1989) 247.}
\lref\ofer{O. Aharony, S. Gubser, J. Maldacena, H. Ooguri and Y. Oz, 
``Large $N$ field theories, string theory and gravity," hep-th/9905111, 
Phys. Rep. {\bf 323} (2000) 183.}
\lref\lp{J.M.F. Labastida and E. P\'erez, ``A relation between the 
Kauffman and the HOMFLY polynomial for torus knots," q-alg/9507031, 
J. Math. Phys. {\bf 37} (1996) 2013.}
\lref\homfly{P. Freyd, D. Yetter, J. Hoste, W.B.R. Lickorish, K. Millet
and 
A. Ocneanu, ``A new polynomial invariant of knots and links," Bull. Amer. 
Math. Soc. {\bf 12} (1985) 239.}
\lref\d{M.R. Douglas, ``Conformal field theory techniques in large $N$ 
Yang-Mills theory," hep-th/9311130.}
\lref\quanta{S.~Elitzur, G.~Moore, A.~Schwimmer and N.~Seiberg,
``Remarks on the canonical quantization of the Chern-Simons-Witten
theory,''
Nucl.\ Phys.\  {\bf B 326} (1989) 108. J.M.F. Labastida 
and A.V. Ramallo, ``Operator formalism for Chern-Simons theories,''
Phys.\ Lett.\  {\bf B 227}(1989) 92; ``Chern-Simons theory and conformal 
blocks,'' Phys.\ Lett.\  {\bf B 228} (1989) 214. S. Axelrod, S. Della
Pietra, 
and E. Witten, ``Geometric quantization of Chern-Simons gauge theory,'' 
J. Diff. Geom. {\bf 33} (1991) 787.} 
\lref\lpp{J.M.F. Labastida and E. P\'erez, ``Gauge-invariant 
operators for singular knots in
Chern-Simons gauge theory," hep-th/9712139, Nucl.\ Phys.\ {\bf
B 527} (1998) 499.}
\lref\witop{E. Witten, ``Chern-Simons gauge theory as 
a string theory,'' hep-th/9207094, in {\it The Floer memorial volume}, 
H. Hofer, C.H. Taubes, A. Weinstein and E. Zehner, eds., 
Birkh\"auser 1995, p. 637.}  
\lref\peri{V. Periwal, ``Topological closed-string theory 
interpretation of Chern-Simons theory,'' hep-th/9305112, Phys. Rev. Lett. 
{\bf 71} (1993) 1295.}
\lref\douglasn{M.R. Douglas, ``Chern-Simons-Witten theory as a topological 
Fermi liquid,'' hep-th/9403119.}
\lref\awada{M.A. Awada, ``The exact equivalence of Chern-Simons 
theory with fermionic string theory,'' Phys. Lett. {\bf B 221} (1989) 21.}
\lref\lm{J.M.F. Labastida and M. Mari\~no, ``Polynomial invariants 
for torus knots and topological strings,'' hep-th/0004196.}
\lref\coleman{S. Coleman, {\it Aspects of symmetry}, Cambridge University 
Press.}
\lref\wzw{S.G. Naculich and H.J. Schnitzer, ``Duality between 
$SU(N)_k$ and $SU(k)_N$ WZW models,'' Nucl. Phys. {\bf B 347} (1990) 687.}
\lref\ns{S.G. Naculich, H.A. Riggs and H.J. Schnitzer, ``Group level 
duality in WZW models and Chern-Simons theory,'' Phys. Lett. {\bf B 246} 
(1990) 417. E.J. Mlawer, S.G. Naculich, H.A. Riggs and H.J. Schnitzer, 
``Group-level duality of WZW coefficients and Chern-Simons link
observable,'' 
Nucl. Phys. {\bf B 352} (1991) 863. 
S.G. Naculich, H.A. Riggs and H.J. Schnitzer, ``Simple-current 
symmetries, rank-level duality, and linear skein relations for
Chern-Simons 
graphs,'' hep-th/9205082, Nucl. Phys. {\bf B 394} (1993) 445.}
\lref\alp{M. \'Alvarez, J.M.F. Labastida, and E. P\'erez, 
``Vassiliev invariants for links from Chern-Simons perturbation theory,'' 
hep-th/9607030, Nucl. Phys. {\bf B 488} (1997) 677.}
\lref\ilm{E. Witten, ``Gauge theories and integrable lattice models,'' 
Nucl. Phys. {\bf B 322} (1989) 629.}
\lref\gmm{E. Guadagnini, M. Martellini and M. Mintchev, ``Wilson 
lines in Chern-Simons theory and link invariants,'' Nucl. Phys. 
{\bf B 330} (1990) 575.}
\lref\gross{D. Gross, ``Two-dimensional QCD as a string theory,'' 
hep-th/9212149, Nucl. Phys. {\bf B 400} (1993) 161.}
\lref\grosstaylor{D. Gross and W. Taylor, ``Two-dimensional QCD is 
a string theory,'' hep-th/9301068, Nucl. Phys. {\bf B 400} (1993) 181. 
``Twists and loops in the string theory of two-dimensional QCD,'' 
hep-th/9303046, Nucl. Phys. {\bf B 403} (1993) 395.}
\lref\dmvv{R. Dijkgraaf, G. Moore, H. Verlinde and E. Verlinde, 
``Elliptic genera of symmetric products and second quantized strings,'' 
hep-th/9608096, Commun. Math. Phys. {\bf 185} (1997) 197. R. Dijkgraaf, 
``Fields, strings, matrices and symmetric products,'' hep-th/9912104, 
in {\it Moduli of curves and abelian varieties}, C. Faber and 
E. Looijenga, eds.,  }
\lref\lickm{W.B.R. Lickorish and K.C. Millett, ``A polynomial invariant of 
oriented links,'' Topology {\bf 26} (1987) 107.}
\lref\nrsym{S.G. Naculich, H.A. Riggs and H.J. Schnitzer, ``The string 
calculation of Wilson loops in two-dimensional Yang-Mills theory,'' 
hep-th/9406100, Int. J. Mod. Phys. {\bf A 10} (1995) 2097.}
\lref\rama{P. Ramadevi and T. Sarkar, ``On link invariants and 
topological string amplitudes,'' hep-th/0009188.}
\lref\rgk{P. Rama Devi, T.R. Govindarajan and R.K. Kaul, 
``Three-dimensional Chern-Simons theory as a theory of knots and 
links (III). Compact semisimple group,'' hep-th/9212110, 
Nucl. Phys. {\bf B 402} (1993) 548.}
\lref\ofer{O. Aharony, S. Gubser, J. Maldacena, H. Ooguri and Y. Oz, 
``Large $N$ field theories, string theory and gravity," hep-th/9905111, 
Phys. Rep. {\bf 323} (2000) 183.}
\lref\milnor{J. Milnor, {\it Singular points of complex hypersurfaces},
 Annals of Mathematics Studies, No. 61, Princeton University Press, 1968.}


\Title{\vbox{\baselineskip12pt
\hbox{US-FT-17/00}
\hbox{RUNHETC-2000-38}
\hbox{HUTP-00/A042}
\hbox{hep-th/0010102}
}}
{\vbox{\centerline{Knots, Links and Branes at Large $N$}
}}
\centerline{J.M.F. Labastida$^{a}$, Marcos Mari\~no$^{b}$ 
and Cumrun Vafa$^{c}$}

\bigskip
\medskip
{\vbox{\centerline{$^{a}$ \sl Departamento de F\'\i sica 
de Part\'\i culas,
Universidade de Santiago de Compostela}
\vskip2pt
\centerline{\sl E-15706 Santiago de Compostela, Spain}}

\bigskip
\medskip
{\vbox{\centerline{$^{b}$ \sl New High Energy Theory Center, 
Rutgers University}
\vskip2pt
\centerline{\sl Piscataway, NJ 08855, USA }}
\bigskip
\medskip
{\vbox{\centerline{$^{c}$ \sl Jefferson Physical 
Laboratory, Harvard 
University}
\vskip2pt
\centerline{\sl Cambridge, MA 02138, USA }}

\bigskip
\bigskip
\noindent
We consider Wilson loop observables for Chern-Simons theory
at large $N$ and its topological string dual
 and extend the previous checks for this duality to the case
of links.
We find an interesting structure involving
representation/spin degeneracy of
branes ending on branes which features in the large $N$
dual description of Chern-Simons theory. This leads to a refinement
of the integer invariants for links and knots. We illustrate 
our results with explicit computations on the Chern-Simons 
side.

\bigskip

\Date{October 12, 2000}


\newsec{Introduction}
The large $N$ limit of Chern-Simons theory on ${\bf S}^3$
has been conjectured
to be equivalent to topological strings on the blown
up conifold geometry \gvtwo.  The implications of this conjecture
for the Wilson loop observables have been formulated in \ov\ 
and checked for the case of the unknot. In particular, this 
conjecture implies that the knot invariants of Chern-Simons theory 
can be formulated in terms of new integer invariants \ov.  
The requisite Wilson loop observables
for torus knots
were computed in \lm\ and found to be in agreement with the 
predictions of this large $N$ conjecture.  The large $N$
conjecture has also been checked for some other knots in \rama .

The structure of the large $N$ result suggests that
arbitrary Wilson loop observables are related on the
closed topological string side, to considering
the propagation of topological strings in the
blown up conifold geometry with a background non-compact
D-brane.   Thus the results of this duality can be turned
around and be used for computations of topological strings
with boundaries--a subject which is not very well understood
at the present.

The aim of this paper is to extend
these results in a number of directions.  First we wish to gain
a deeper understanding of the structure
of the new integer invariants.  In particular the results
of \lm\ shows that there are some universal regularities
in these numbers that requires explanation.  We
explain these universal properties in this paper by
studying the problem of counting degeneracy of D2 branes
ending on D4 branes.  Moreover this leads to 
a refinement of the basic integer invariants introduced 
in \ov. Secondly we wish to extend these results
to links.  This involves, on the Chern-Simons computational
side, a careful treatment of link invariants consistent
with the large $N$ expansion.  The results are found
to agree with predictions of the large $N$ duality.
  Thirdly, we construct
the relevant Lagrangian D-brane on the topological string side, 
for a large class of algebraic knots.

The organization of this paper is as follows:
In section 2 we review the large $N$ Chern-Simons/topological
string duality, including the incorporation of Wilson
loop observables.
In section 3 we analyze the BPS structure of D2 branes
ending on D4 branes which leads us to a refinement of
the integer invariants associated to knots and links.
In section 4 we formulate the link invariants in the context
of large $N$ Chern-Simons theory.  In section 5 we construct
the relevant Lagrangian submanifold associated to a large
class of algebraic knots, needed for the closed string
dual. Finally, in section 6 we present many explicit results 
for the refined integer invariants for some knots and links. 

\newsec{Chern-Simons theory and topological strings}

It was conjectured in \gvtwo\ that $SU(N)$ Chern-Simons gauge theory
on $S^3$ at large $N$ is equivalent to topological strings on
the non-compact Calabi-Yau threefold, with local geometry
given by the ${\cal O}(-1)+{\cal O}(-1)$ bundle over ${\bf P}^1$.  This 
conjecture was motivated by considering topological D3-branes
for which the worldvolume theory gives rise to $SU(N)$ Chern-Simons
theory \witop .  Namely one considers the conifold geometry
$$z_1^2+z_2^2+z_3^2+z_4^2=\mu$$
where $z_i$ are complex parameters.  If we decompose
the complex coordinates to real coordinates,
i.e., if we write
$$z_j=x_j+ip^j$$
then one can see that
the above equation has the geometry of $T^* {\bf S}^3$ where ${\bf S}^3$
is given by $p^i=0$, and the $p^i$ denote the cotangent directions.
If we consider $N$ topological $D3$ branes wrapped on the ${\bf S}^3$,
then
on ${\bf S}^3$ we get an effective string theory which is an $SU(N)$ 
Chern-Simons theory \witop, at level $k$,
 where the string coupling is $g_s=2\pi i /(k+N)$
(the shift in the denominator $k\rightarrow k+N$
 is the familiar quantum correction
in the context of CS theory). The action is
\eqn\csaction{
S={k \over 4\pi} \int_M {\rm Tr} \Bigl( A\wedge d A + {2 \over 3} A
\wedge A \wedge A \Bigr),} 
where $A$ is a gauge connection on some $SU(N)$ vector bundle over a
 three-manifold $M$, which in this case we take to
 be ${\bf S}^3$. 

The conjecture of \gvtwo\ is that at
large $N$ the D-branes disappear, and instead
the conifold undergoes a transition where
${\bf S}^3$ shrinks to zero size ($\mu \rightarrow 0$) and a ${\bf P}^1$
grows
whose K\"ahler parameter $t$ is given by
$$t= N g_s={2\pi i N\over k+N}$$
This conjecture has been checked for the free energy on both sides
 to all orders in the $1/N$ expansion.  The answer 
for the partition function on the Chern-Simons
 side was derived in \cs\ and on the closed string side the topological
string amplitude was computed using M-theory \gv\ and 
also using the mathematical definition of topological strings
in \fp.  This duality has been recently
been embedded in the context of type IIA superstrings
with $N=1$ supersymmetry in 4 dimensions \vare\ 
and it provides
a new context where a background involving a large number of $D$-branes
is equivalent to another background without any $D$-branes.

There is more to the Chern-Simons theory than just the partition function.
In particular there is a rich set of observables associated
to links in ${\bf S}^3$ and representations of the gauge group $SU(N)$ 
\cs.
In particular, 
from the holonomy of 
the gauge field around a closed loop $\gamma$ in $M$,
\eqn\holo{
U={\rm P}\,\exp\, \oint_{\gamma} A,}
one can construct a natural class of observables, the 
gauge-invariant Wilson loop operators, which are given by
\eqn\wilso{
W^{\gamma}_R(A)={\rm Tr}_R \, U,}
where $R$ denotes an irreducible representation of $SU(N)$.
Some of the standard topological invariants that have been considered in
the
context of Chern-Simons gauge theory are vevs of products of these
operators:
\eqn\correlat{
\langle W^{\gamma_1}_{R_1}\cdots W^{\gamma_n}_{R_n}\rangle =
{1\over Z(M)}\int [{\cal D} A] \Bigl( \prod_{i=1}^n W_{R_i}^{\gamma_i} 
\Bigr) {\rm e}^{iS},}
where $Z(M)$ is the partition function of the theory. These vevs 
are functions of the variables
\eqn\variar{
q={\rm e}^{g_s}=\exp \Bigl[ {2 \pi i \over k +N} \Bigr],\,\,\,\,\,\ 
\lambda^{-1}=q^{-N}={\rm e}^{-t}.} 

In the context of topological strings, i.e., with $N$ D3 branes
on ${\bf S}^3$ embedded in the conifold background,
 it is natural to ask
how these observables can arise.  This was answered in \ov :
Consider a number of $\gamma_{\alpha} \in {\bf S}^3$, $
\alpha=1, \cdots, L$.  Then to each one of them
one associates a Lagrangian submanifold $D_\alpha$ which intersects
${\bf S}^3$ on $\gamma_{\alpha}$.  This is obtained by appending to each
point of $\gamma_{\alpha}$ a two dimensional subspace ${\bf R}^2$ which is
the set of momentum vectors $p$ orthogonal to the velocity
vector $d\gamma_{\alpha}/ds$.  Note that the topology of each $D_\alpha$
is ${\bf R}^2\times {\bf S}^1$.
 We consider placing $M_{\alpha}$
branes on the cycle $D_\alpha$.  Since
these are non-compact we can consider them
as non-dynamical.  In particular on each of them lives an
$SU(M_{\alpha})$ gauge fields which we consider as classical.
Let $V_{\alpha}$ denote the holonomy of this gauge field
along the $\gamma_{\alpha} \in D_{\alpha}$.  As far as the
Chern-Simons gauge theory on ${\bf S}^3$ is concerned we get additional
massless
fields living on the $\gamma_\alpha$ corresponding
to open strings stretched between branes on $D_{\alpha}$ and ${\bf S}^3$.
Integrating them out leads (as discussed in \ov ) to the operator
\eqn\gen{
Z(U_\alpha,V_\alpha)=\exp\bigl[
\sum_{\alpha=1}^L \sum_{n=1}^\infty {1 \over n} {\rm Tr}\, U_{\alpha}^n\, 
{\rm Tr }\, V_{\alpha}^n\bigr],}
 In this
operator the trace is taken over the fundamental representation 
\foot{In what follows,
when no representation is indicated in a trace, it should be understood
that it
must be taken in the fundamental representation.}. 

If one can evaluate
the expectation value $\langle 
Z(U_\alpha,V_\alpha) \rangle= {\rm exp}(F(V_\alpha))$ for all links, 
then effectively one can compute all the topological observables
of Chern-Simons theory \correlat\ by rewriting
the holonomy in representation $R$, ${\rm Tr}_R U$ in terms of the trace
of powers of holonomy in fundamental representation, appearing in
$Z(U_{\alpha},V_{\alpha})$.

It is natural then to ask how these correlators behave at large $N$.
Since according to the conjecture \gvtwo\ we should
be effectively ending up with topological
strings on the blown up conifold geometry where the D-branes
wrapped around ${\bf S}^3$ have disappeared, we need to know what
happens to the other D-branes wrapped over the
non-compact $D_\alpha$ cycles
after this transition. Geometrically it is clear that they
cannot disappear (as the only thing that shrinks is the ${\bf S}^3$)
and so they should manifest themselves as some Lagrangian
submanifolds on the other side. Let us continue to denote
the corresponding Lagrangian submanifold by $D_\alpha$. Moreover, there
should be a distinguished ${\bf S}^1$ on the Lagrangian submanifold
after the transition, for which we consider the holonomy $V_{\alpha}$.
 Thus the statement of the conjecture
would be that $F(V_\alpha)$ can be computed by considering
the topological strings on the blown up geometry with extra
D-branes $D_\alpha$.  In particular
\eqn\mini{
F(V_{\alpha})=\sum_{g=0}^\infty 
\sum_{h=1}^\infty \sum_{\alpha, n_1, \cdots, n_h} 
g_s^{2g-2 +h}F^{\alpha}_{g;n_1, \cdots, n_h}\, {\rm
Tr}\,V_{\alpha_1}^{n_1} 
\cdots {\rm Tr}\, V_{\alpha_h}^{n_h},} 
where $F^{\alpha}_{g;n_1, \cdots, n_h}$ denotes the topological
string amplitude on the blown up geometry with genus $g$ and with
$h$ holes, where the $i$-th hole is mapped to the $D_{\alpha_i}$ brane
and winds around the corresponding circle $n_i$ times.  The computation
of these amplitudes is very difficult, and no techniques are currently
available to do a direct computation.  However, some general
features of topological amplitudes can be deduced.  This was
done in \ov\ following the idea of \gv\gvm , which relates the degeneracy 
of D2 branes, including both their charge (wrapping
around 2-cycles) and spin, 
to the topological string amplitudes.  The novel feature in the
case at hand is that one has to consider branes ending on the
Lagrangian submanifold.  Moreover, in addition to the
``bulk'' charge and spin, the D2 brane also forms
a representation of $SU(M)$.  It was shown in \ov\ that 
by relating the
topological string amplitudes to type IIA superstring amplitudes 
in 2 dimensions with background D4 branes, and considering
the contribution of D2 branes ending on D4 branes to topological
string amplitudes,
one can deduce
the following structure for the partition function:
\eqn\conj{
F(V)  =\sum_{d=1}^\infty \sum_{R} f_R (q^d, \lambda^d) {\rm Tr}_R 
{V^d\over d} ,}
where
\eqn\ovconj{ 
f_R(q, \lambda)=\sum_{s,Q}{ N_{R,Q,s}\over q^{{1\over 2}}-q^{-{1\over 2}}}
 \lambda^Q q^s .}
Here, the $N_{R,Q,s}$ 
are integers and denote the degeneracy
of D2 branes of bulk charge $Q$ with 2d spacetime spin $s$ and $SU(M)$
representation $R$. The $Q$, for a given representation
$R$, denotes an
element of the relative homology $H_2(X,D)$
where $X$ is the CY ${\cal O}(-1)+{\cal O}(-1)$ over ${\bf P}^1$ and
$D$ is the Lagrangian submanifold.  The difference of two
$Q$'s is an integer for any fixed $R$.  In the examples
we will deal with it turns out that $Q$'s are integer
or half-integer.  
 Here we have written
the partition function for the case of a single 
D-brane, but that can be easily generalized.
The integrality properties of $N_{R,Q,s}$ are rather
non-trivial and this gives a strong constraint on any
proposed answer for the topological string amplitudes
involving D-branes.

To check the large $N$ conjecture for the Wilson loop observables
one has to overcome a number of obstacles.  On
the gauge theory side we have to compute
the Chern-Simons observables for arbitrary links and representations.
On the gravity side (i.e., after the transition) we need
to construct
the $D_\alpha$'s.  Finally we have to find
a way to compute the topological string amplitudes 
in the presence of these D-branes, which is
equivalent
to computing the degeneracy numbers 
$N_{R,Q,s}$.\foot{By analytic continuation one can also
express the topological string amplitudes in terms of
the conjugate quantum numbers by flipping the sign, i.e.,
$N_{R^*,-Q,-s}=-N_{R,Q,s}$, where $R^*$ denotes the conjugate
representation.}

These questions were answered in \ov\ for a very simple
observable, namely the unknot, which is just a single
$\gamma$ which is not knotted.  The Lagrangian submanifold
after the transition side was identified for this case
(by identifying it as the invariant locus of an anti-holomorphic
involution). In particular it intersected the ${\bf P}^1$ on
an equator.
It was shown in that case that the
geometry of D2 branes is particularly simple and there are only
two D2 branes ending on the Lagrangian submanifold:  
one corresponding to the D2 brane wrapped over the northern
hemisphere, and the other over the southern hemisphere.  These
have $Q$ charge $\pm 1/2$ and spin $s=0$, and are in the fundamental
representation, namely
$N_{\tableau{1},\pm 1/2 ,0}=\pm 1$ and the rest
of the $N_{R,Q,s}$ are zero.

To further test the above predictions at large $N$,
more interesting knots were considered in \lm . 
In particular a class of knots known as $(n,m)$
torus knots were studied. Already the Chern-Simons
computation is very non-trivial for these knots, as one
has to find a way to compute arbitrary number
of product of the trace of the holonomy observable in
{\it all} representations of $SU(N)$.  Moreover one
has to relate this with the particular combinations
of observables given in \conj\ovconj.  For this comparison
it turns out natural to reexpress the Chern-Simons observables
in terms of powers of $U$ in the fundamental
representation. In particular
if $R$ is a representation of $SU(N)$ with $\ell$ boxes
then ${\rm Tr}_R U$ is a universal polynomial in 
${\rm Tr} U^k$ in the fundamental representation of total
degree $\ell$. Then the question of computation gets
transformed to the computation of product of various
powers of ${\rm Tr} U^k$.  At the end of the day, what one
finds is that if one knows the expectation value
of the single $\langle {\rm Tr}_R U \rangle $ for any representation
$R$ of $SU(N)$ with up to $\ell$ boxes, one can find 
$f_{R'}$ for representation $R'$ of $SU(M)$ with up to
$\ell$ boxes.  This gives a recursive way
to compute the $f_{R'}$ organized in increasing
order in terms of the
number of boxes.  Let us review this in a little
more detail.

As shown in \lm, to compare the large $N$ topological
string predictions \conj\ovconj\ with the Chern-Simons
computation, it is convenient to introduce the 
following basis on the space of Wilson loop operators:  
\eqn\basis{
\Upsilon_{\vec k}(U)= \prod_{j=1}^\infty \Big( {\rm Tr}\,U^j \Big)^{k_j}, 
}
which are labeled 
by a vector $\vec k$ of nonnegative integers, and $U$ is the holonomy
\holo. Given such a vector, 
we define:
\eqn\long{
\ell=\sum j \,k_j,\,\,\,\,\,\ |\vec k| =\sum k_j.} 
We can associate to any vector $\vec k$ a conjugacy class $C(\vec k)$ of
the  
permutation group  $S_\ell$. This class has $k_1$ cycles of 
length 1, $k_2$ cycles of length 2, and so on. The number of elements 
of the permutation group in such a 
class is given by \fh\
\eqn\conjug{
|C(\vec k)|= {\ell! \over \prod k_j! \prod j^{k_j}}.}
Notice that the vectors $\vec k$ with $\sum_j\, j k_j=\ell$ 
are in one-to-one 
correspondence with the partitions of $\ell$. We also define 
generalized connected vevs as follows: 
first, associate to any $\vec k$ the polynomial $p_{\vec k}(x)=
 \prod_j x_j^{k_j}$ 
in the variables $x_1, x_2, \dots$. We then define the
 ``connected" coefficients 
$a_{\vec k}^{(c)}$: 
\eqn\cumul{
\log\Bigg(1+ \sum_{\vec k} {|C(\vec k)| \over \ell!} a_{\vec k}p_{\vec
k}(x) \Bigg)= 
\sum_{\vec k} {|C(\vec k)| \over \ell!} a^{(c)}_{\vec k}p_{\vec k}(x).}
We also introduce the following notation for the vevs of the operators 
in \basis:
\eqn\vevs{
G_{\vec k} = \langle \Upsilon_{\vec k} (U) \rangle.} 
Using \conj\ovconj\ one deduces the following relation \lm:
\eqn\master{
G_{\vec k}^{(c)}(U)= \sum_{n|\vec k} n^{|\vec k| -1}
 \sum_R \chi_R (C(\vec k_{1/n}))f_R(q^n, \lambda^n),} 
where $\chi_R( C(\vec k))$ denotes the character of the representation $R$ 
of the symmetric group for the conjugacy class $C(\vec k)$ specified by 
the vector $\vec k$.  
In this equation, the vector $\vec k_{1/n}$ is defined as 
follows. Fix a vector $\vec k$, and 
consider all the positive integers that satisfy the following condition:
$n|j$ for every $j$ with $k_j \not=0$. When this happens, 
we will say that ``$n$ divides 
$\vec k$,'' and we will denote this as $n|\vec k$.  
We can then define the vector $\vec k_{1/n}$ whose components are:
\eqn\shifted{
(\vec k_{1/n})_i=k_{ni}.}
The vectors which satisfy the above condition and are
 ``divisible by $n$'' have the structure 
$(0,\dots, k_n, 0,\dots,0, k_{2n},\dots)$, and the vector 
$\vec k_{1/n}$ is then
given by $(k_n,k_{2n},\dots)$.

One can extract the generating functions $f_R(q, \lambda)$ 
from the connected vevs in Chern-Simons theory, using the 
relation \master. This was done in \lm\ to 
check, for the torus knots, the structure predicted in  \ov\ 
for the large $N$ behavior of the Wilson loop
observables.  In particular the integrality properties
of the topological string amplitudes were verified. The results 
of \lm\ strongly suggest
that for non-trivial torus
knots, the integer invariants 
$N_{R,Q,s}$ are non-vanishing for {\it all} representations.
Moreover for each representation $R$, there are a finite number of
values of $Q,s$ for which the $N$'s are non-vanishing. These
$N_{R,Q,s}$ would give the number of D2 branes in the blown
up geometry which end on the D-branes wrapping around the
Lagrangian submanifold corresponding to torus knots
after the transition. These Lagrangian submanifolds
were not constructed in \lm , but we will present
a proposal for them in section 5 of this paper.  Even
knowing the Lagrangian submanifold is not sufficient
to give the integer invariants, since there are no known techniques
in complicated enough cases as we are encountering
to directly compute $N_{R,Q,s}$'s.  So this large
$N$ computation of Chern-Simons knot invariants
should be viewed as a powerful
technique for computing degeneracies of branes
in this geometry.

Even though all the predicted aspect of the large
$N$ theory were checked for torus knots in \lm ,
 more structure was found in \lm\ which needed
further explanation. In particular it was noted there that 
the topological string answer predicts certain surprising universal
relations
among the BPS degeneracies. For example a simple (inductive) reformulation
of the results in \lm\ shows that
\eqn\nrel{
N_{R,Q,s}=(-1)^{\ell -1}N_{R^t,Q, -s}.}
where $R^t$ denotes the $SU(M)$ representation which has
the transposed Young tableau of the one corresponding to $R$, 
and $\ell$ is the number of boxes in the Young tableau.
In fact this follows by checking the agreement
of the large $N$
expansion encoded in terms of the topological closed string amplitude
\mini\ with that given in \conj\ovconj. To see this notice that, after 
performing an analytic continuation to a series involving only positive 
$n_i$, \mini\ can be written as
\eqn\frek{
F(V)= \sum_{\vec k} 
\sum_{g=0}^{\infty} F_{g, \vec k}(\lambda) g_s^{2g-2+ |\vec k|} 
\Upsilon_{\vec k} (V),} 
where $k_i$ is the number of $i$'s in the $h$-uple $(n_1, \cdots, n_h)$, 
so that $|\vec k|=h$. In the context of Chern-Simons theory, 
\frek\ is nothing but the usual $1/N$ expansion of the 
connected vevs of Wilson loops. Note that the parity of the power of 
$g_s$ in \frek\ correlates with the number of holes, and translating this
to 
\conj\ovconj\ using Frobenius relation and 
\eqn\symchar{
\chi_{R^t}(C(\vec k))=(-1)^{|\vec k| + \ell} \chi_R(C(\vec k)),} 
yields the above relation
among $N_{R,Q,s}$'s. There
were also additional vanishing relations noted in \lm\ from 
comparing \frek\ and \conj\ovconj\ which were also
in need of explanation. In this paper we will be able to explain
all those relations (using some plausibility assumptions) by
studying more carefully the degeneracy of D2 branes ending
on D4 branes 
and in particular we will have a reformulation
of $N_{R,Q,s}$ in terms of other more basic integer quantities 
${\widehat N}_{R,Q,g}$ which are not restricted. 

Further checks for the large $N$ predictions
of Wilson loop observable, involving 
certain knots up to nine crossings, were recently done in \rama\ 
using the techniques of \rgk. 

\newsec{Degeneracy of D2-branes ending on D4-branes}

As discussed before, there are hints that the numbers $N_{R,Q,s}$
satisfy certain universal relations.  We wish to derive this
fact from the viewpoint of counting the degeneracy of D2 branes
ending on D4 branes in the stringy realization of topological
strings with D-branes.

Consider type IIA strings on a Calabi-Yau threefold (which for most of
the applications we have in mind will be non-compact). Suppose
we have $M$ D4-branes wrapping a 3-dimensional Lagrangian submanifold
of the Calabi-Yau, and filling a 2-dimensional spacetime subspace of 
${\bf R}^4$.
Suppose also that for the Lagrangian submanifold we have $b_1=1$. 
As discussed in \ov\ this gives rise, in the ${\bf R}^2$ subspace, to a
$U(1)^M$ magnetic gauge theory.
We will be interested in the degeneracy of D2 branes ending on D4 branes
in this geometry. These can be viewed as having some $U(1)^M$ charge. 
 By the $S_M$ permutation invariance of the D4 branes,
it follows that the particles will form representations of $U(1)^M/S_M$,
which can also be viewed as linear combination of representations of
$U(M)$.
Let us label such a representation with $R$.  The D2-brane will also
have some charge $Q$ corresponding to which 2-cycle (in
the sense of relative homology) of the Calabi-Yau
it wraps around.  It also has some $SO(2)$ spin $S$.  To define
this more precisely, we have to consider the M-theory description
of this geometry \ov .  This means
that we consider M-theory on CY 3-fold
with M5 branes filling ${\bf R}^3$ and wrapping over the
Lagrangian submanifold in the CY.  We relate the $SO(2)$ spin $S$
 to the 3-dimensional rotation
symmetry on the non-compact
worldvolume of the M5 brane.  
There is also an $SO(2)$ R-symmetry for 4 supercharges
in $d=3$.  For BPS states a combination of supercharges which 
is neutral under $S_L=S+R$ annihilates the state, while
the other supercharge which changes $S_L$, but is neutral
under $S_R=S-R$, generates the BPS multiplet. 
This is very similar to the BPS structure for a theory
in 5 dimensions with 8 supercharges (such as CY 3-fold compactifications
of M-theory).  In that case one has the rotation group $SO(4)=
SU(2)_L\times SU(2)_R$ and the active supercharge
of the BPS states resides in one of the $SU(2)$'s.  This
was in fact strongly used in relating the degeneracy
of D2 branes wrapped around cycles of CY 3-fold with the
topological string amplitudes \gvm .
The $S_L$ and $S_R$ of the 3d theory we have been discussing
correspond to the $J^3_L$ and $J^3_R$ of these $SU(2)$'s respectively.
In fact the presence of the M5 brane breaks the $SO(4)$ rotation
symmetry of 5d theory to $SO(2)_S\times SO(2)_R$ which gives us
the above identification of $S_{L,R}$ with $J^{3}_{L,R}$.

The number of BPS states may change in general, if two
short multiplets combine to a bigger multiplet.  There
is a way to define an index which is the net number of
BPS states which cannot recombine.  Namely, if we take a trace
over all BPS states of a given charge and representation quantum numbers, 
as well as a fixed $S_L$,
weighted with ${\rm Tr} (-1)^{2S_R}$, one gets zero for pairs of short
multiplets that can combine to a long multiplet. Thus
this combination is invariant under deformation.  The number $N_{R,Q,s}$
refers to this net number where $s=S_L$.

We are now ready to study the net degeneracy of D2-branes ending
on D4-branes, in the sense defined above. Similar
to what was done in the context of the closed string case \gvm\
we first consider the situation where we have a D2-brane
of a fixed genus $g$ with $\ell$ holes ending on the D4-brane.  We
will assume for simplicity that if there are moduli for these D2-branes
the genus and the number of holes does not change over all of this
moduli space.  Let us call this moduli space ${\cal M}_{g,\ell}$.
We will take this to be the moduli space with $\ell$
{\it ordered } holes (later we will divide by the action of
permutation group $S_\ell$ to find the physical number of states).
 
As discussed in \gvm\ we have to study the moduli space
to D2-branes ending on D4-branes together with a flat
bundle on it.  For a genus $g$ surface with $\ell$ holes this
gives rise to the Jacobian $J_{g,\ell}={\bf T}^{2g+\ell-1}$.  The
assumption that there is no degeneration of the Riemann
surface along the moduli, means
that the Jacobian is always a constant ${\bf T}^{2g+\ell -1}$ (otherwise
we would have complications similar to the ones discussed in 
\gvm\kkv\ in the context of closed strings).  
There is only one
more ingredient:  for each boundary we should decide which
of the $M$ branes it lives on.  Thus we naturally get, for the
Hilbert space, a tensor product of $F^{\otimes \ell}$, where $F$
is the fundamental representation of $SU(M)$.
 Therefore, the Hilbert space 
associated naturally to the above geometrical configuration is given by 
\eqn\conf{
F^{\otimes \ell} \otimes H^*(J_{g,\ell}) 
\otimes H^*({\cal M}_{g,\ell}).} 
An important 
point is that this Hilbert space is associated with the
moduli space of $\ell$ {\it distinguished} holes, which is not
physical, and we have to mod out by the action of the permutation
group $S_{\ell}$.
We can factor out the cohomology of the Jacobian ${\bf T}^{2g}$
of the ``bulk'' 
Riemann surface, $H^*({\bf T}^{2g})$, since the permutation group 
does not act on it. The projection onto the symmetric piece can 
be easily done using the Clebsch-Gordon coefficients
$C_{R\,R'\,R"}$ of the permutation group $S_{\ell}$:    
\eqn\hilbert{
\eqalign{
&{\rm Sym}\Bigl(F^{\otimes \ell} \otimes H^*(({\bf S}^1)^{\ell-1}) 
\otimes H^*({\cal M}_{g,\ell})\Bigr)=\cr &
\,\,\,\,\,\,\ \sum_{R\,R'\,R''} C_{R\,R' \, R''}
{\bf S}_R(F^{\otimes \ell})\otimes{\bf S}_{R'}(H^*(({\bf S}^1)^{\ell-1})) 
\otimes {\bf S}_{R''}(H^*({\cal M}_{g,\ell})),\cr}
 }
where the subscripts of the representation of the vector spaces
mean projecting the space to the corresponding subspace. 
The space ${\bf S}_R(F^{\otimes \ell})$ is nothing but the 
vector space underlying the irreducible representation $R$ of $SU(M)$.
In other words, even though to begin with $R$ labels representations of
$S_{\ell}$ the projection also leads to a definite representation
of the $SU(M)$ group, due to the relation between representation of
symmetric groups and that of $SU(M)$.  So in this sense the same Young
tableau will denote also the representation content of $SU(M)$.
We will thus use the label $R$ in both senses. 

We have to be more specific about the action of the permutation group 
on the cohomology elements. Let's denote by $H=H^*({\bf S}^1)$ the 
cohomology of the circle.   
Although the permutation group $S_{\ell}$ acts in a natural way 
on a Riemann surface with $\ell$ boundaries, there are only $\ell-1$ 
independent one-forms associated to the boundary. This is 
because the one-forms $d\theta_i$, $i=1, \cdots, \ell$, which are 
Poincar\'e dual to the holes in the Riemann surface, 
satisfy $\sum_i d\theta_i=0$. The procedure to 
construct the Hilbert space ${\bf S}_R(H^{\ell -1 })$ is then 
as follows. We consider the Hilbert space 
$H^{\ell}$ generated by 
$\ell$ fermion fields $\psi_i$, $i=1, \cdots, \ell$ 
acting on the vacuum $|0\rangle$ and we decompose it with respect to the 
different representations $R$ by using the Young symmetrizers 
of the corresponding tableaux, and taking into 
account the Grassmann nature of the fermions. 
Finally, we impose the linear constraint $\sum_i \psi_i=0$. It is
instructive 
to consider the simple case $\ell=2$ in some detail. This would for
example
arise for the D2-brane with the topology of annulus.
The space $H^{2}$ is spanned by the four 
states $|0\rangle$, 
 $\psi_{1,2}|0\rangle$ and $\psi_1 \psi_2 |0\rangle$. 
The relevant permutation group $S_2$ corresponds to permuting
$\psi_1 \leftrightarrow \psi_2$.
 Projecting 
onto the symmetric and antisymmetric subspaces, we find:
\eqn\symant{
\tableau{2}: \,\,\ |0\rangle, \,\ (\psi_1 + \psi_2)|0\rangle 
\,\,\,\,\ ; \,\,\,\,\ \tableau{1 1}: \,\,\ (\psi_1 - \psi_2)|0\rangle, \,\ 
\psi_1\psi_2|0\rangle.}
Using that $\psi_1 + \psi_2 =0$, the spectrum turns out to be:
\eqn\eltwo{
\tableau{2}: \,\,\ |0\rangle, \,\,\,\,\ \tableau{1 1}: \,\,\
\psi_1|0\rangle.}
To assign spins to these states, first we note
from the discussion above that the relevant notion of spin
is $S_L$, and with respect to $S_L$ the fermions
associated to the choice of the flat connection on Riemann
surface carry spin $1/2$. In other words,
$\psi_i$ have spin $1/2$.  However this does not fix the spin
assignment of all the states, as we have to choose a spin
for the ground state.  It is natural to choose the spin
for the ground state in such a way that in a given multiplet
the average spin is zero.  This is what we will assume (and
is consistent with the choice which naturally arises from the
duality with knots in Chern-Simons theory).  In the case at hand
the two states differ in spin by $1/2$, and so the symmetric choice
of spin assignment is spin $\mp 1/4$.
With our choice of normalization we call this $s=\pm \half$. 
The same procedure gives for $\ell=3$:
\eqn\elthree{
\tableau{3}: \,\,\ |0\rangle, \,\,\,\,\ \tableau{2 1}: \,\,\ (\psi_1 
+ \psi_2)|0\rangle, \,\,\,\,\ \tableau{1 1 1}: \,\,\ \psi_1\psi_2|0\rangle
,}
with spin assignments $-1/2$, $0$, $1/2$ i.e., $s=-1$, $0$, $1$.

It is clear how to proceed to find the spin content of various
representation that arise in this way, when we have more holes. The
symmetric spectrum, corresponding to a Young tableau with $\ell$ boxes 
and only one row, is given by the vacuum $|0\rangle$, and we assign 
it spin $-(\ell-1)/2$. Let's now consider the states that are obtained 
acting with one fermion $\psi_i$ on the vacuum. There are $\ell$ of them, 
forming a reducible representation $(\ell)$ of $S_{\ell}$ which 
decomposes in reducible representations as follows:
\eqn\simpledec{
(\ell)=(\ell -1) \oplus (1).}
The first summand corresponds to the standard representation $V$
 of $S_{\ell}$, 
with a Young tableau of the form:
\eqn\stab{
\tableau{8 1}}
with $\ell-1$ boxes in the first row. 
The second summand corresponds to the trivial representation 
generated by $( \sum_i \psi_i) |0\rangle$, which we are putting 
to zero. To generate the rest of the spectrum, we have just to 
take the antisymmetrized tensor products $\wedge^d \, V$ (since 
$V$ is fermionic). These are irreducible representations of 
$S_{\ell}$ and are called hook representations, since
their Young tableau is of the form
\eqn\hook{
\tableau{6 1 1 1 1}} 
with $\ell-d$ boxes in the first row. We have then obtained 
the spin/representation content of the spectrum: 
the Hilbert spaces ${\bf S}_R(H^{\ell-1})$ are nonempty only for 
hook representations of the form \hook, and in this case 
they contain one state of statistics $(-1)^d$ and total spin 
$-(\ell-1)/2 + d$, which is  equal to the
spin of the vacuum plus $d$ units of the $d$ fermionic fields 
that appear in $\wedge^d \, V$.  
 
It is useful to encode the spectrum associated 
to a Hilbert space ${\cal H}$ in a 
generating function ${\rm Tr}_{\cal H}(-1)^F q^s$. 
The degeneracy and the spin content of the
contribution of a Riemann surface with $\ell$ boundaries
in representation $R$ is summarized in the 
generating function:
\eqn\srq{
S_R(q)={\rm Tr}_{{\bf S}_R(H^{\ell -1})}(-1)^F q^s.}
In fact that is precisely the contribution of the boundary states to
$f_R$. As we have argued, the generating function corresponding 
to the hook representation $R$ is given by,  
\eqn\expsr{
S_R (q)=(-1)^d q^{ -{\ell -1 \over 2}+d} ,}
and $S_R(q)=0$ for the rest of the representations. 
For example for the case of two holes one has 
that $S_{\tableau{2}}(q) =q^{-1/2}$ and 
$S_{\tableau{1 1}}(q)=-q^{1/2}$, while for $\ell=3$ one has
\eqn\lthree{
S_{\tableau{3}}(q)=q^{-1}, \,\,\ S_{\tableau{2 1}}(q)=-1, \,\,\ 
S_{\tableau{1 1 1}}(q)=q,}
in agreement with \eltwo\ and \elthree. 

The generating functions $S_R(q)$ have two properties that will be 
needed later. First of all, they satisfy 
\eqn\lrschur{
S_R (q^{-1}) = (-1)^{\ell-1}S_{R^t}(q),}
which is a direct consequence of the explicit expression \expsr. 
The second property that they have is the following. Define 
the following polynomials, which are labeled by a 
conjugacy class $C(\vec k)$ of $S_{\ell}$: 
\eqn\pk{
P_{\vec k}(q)=\sum_R \chi_R(C(\vec k)) S_R(q),} 
which can be understood as the graded trace of 
the element $C(\vec k)$ on the total Hilbert 
space $H^{\ell-1}$:
\eqn\schurpol{
{\rm Tr}_{H^{\ell -1}} (-1)^F C({\vec k}) \, 
q^s.}
If $C(\vec k)$ is the trivial permutation, we just get 
the generating function for $H^{\ell-1}$:
\eqn\total{
q^{-{\ell -1 \over 2}} (1-q)^{\ell-1}=
{  (q^{-{1\over 2}} -q^{1\over 2})^{\ell } 
\over q^{-{1\over 2}} -q^{1\over 2}}
.}
For nontrivial conjugacy classes, with $k_1$ one-cycles, 
$k_2$ 2-cycles, and so on, the above trace can be computed 
on $H^{\ell}$ by grouping the fermions as indicated by 
the conjugacy class: 
\eqn\group{
\psi_{\mu_1} \cdots \psi_{\mu_{k_1}} 
\psi_{\mu_1\, \nu_1} \cdots \psi_{\mu_{k_2}\, \nu_{k_2}} \cdots,}
where $\psi_{\mu_i \, \nu_i \, \cdots}=\psi_{\mu_i} \psi_{\nu_i} 
\cdots$ are ``composites'' which for a cycle of length $j$ have 
spin $j$, and there are $k_j$ of them. It is then easy to show that
\eqn\expk{
P_{\vec k}(q)={ \prod_j (q^{-{j\over 2}} -q^{j\over 2})^{k_j} 
\over q^{-{1\over 2}} -q^{1\over 2}},} where we have factored 
out a $q^{-{1\over 2}}- q^{1 \over 2}$ which comes from 
imposing the constraint $\sum_i \psi_i=0$. This expression 
will be useful later on. 

We can now give the expression of $f_{R} (q, \lambda)$ 
in terms 
of the structure we have analyzed. 
Note that since the cohomology of ${\cal M}_{g, \ell}$
is represented by fermions which only carry ``right'' spin
$S_R=S-R$ \gvm\ 
we do not need to know their actual cohomology degree in connection with
the topological amplitudes which are only sensitive to $S_L=S+R$. 
Define the integers
\eqn\ints{
{\widehat N}_{R,g,Q}=\chi 
\bigl( {\bf S}_{R}(H^*({\cal M}_{g,\ell}))\bigr).} 
Then collecting the results from our discussion
above, we have
\eqn\fr{
f_{R}(q, \lambda)=
\sum_{g\ge 0} \sum_{Q}
\sum_{R', R''} 
C_{R\,R'\,R''}S_{R'}(q) 
{\widehat N}_{R'',g,Q} 
 (q^{-{1\over 2}}-q^{1\over 2})^{2g-1}\lambda^Q,}
where $(q^{-{1\over 2}}-q^{1\over 2})^{2g}$ comes from the bulk of the 
Riemann surface, as in \gvm, and the extra $(q^{-{1\over 2}}-
q^{1\over 2})^{-1}$ comes from the Schwinger computation \ov. 
It is useful to introduce the generating functions:
\eqn\tildefr{
{\widehat f}_{R}(q, \lambda)= 
\sum_{g \ge 0}\sum_Q {\widehat N}_{R,g,Q}
(q^{-{1\over 2}}-q^{1 \over 2})^{2g-1}\lambda^Q.} We then have 
the relation:
\eqn\relafs{ 
 f_{R}(q, \lambda)=\sum_{R'} 
M_{R\, R'}(q) 
{\widehat f}_{R}(q, \lambda),}
where the matrix $M_{R\, R'}(q)$ is given by
\eqn\matr{
M_{R\,R'}(q)=
\sum_{ R''}
C_{R\,R'\,R''}S_{R''}(q)=
\sum_{{\vec k}} 
 {|C({\vec k})| \over \ell!} 
\chi_{R}(C({\vec k})) \chi_{R'}
(C({\vec k}))P_{{\vec k}}(q).} 
To obtain this expression, we have taken into 
account that the Clebsch-Gordon coefficients are given by: 
\eqn\coefchar{
C_{R\,R'\,R''} =\sum_{\vec k} {|C(\vec k)| \over \ell !} 
\chi_R (C(\vec k))  \chi_{R'} (C(\vec k)) \chi_{R''} (C(\vec k)) .}
The matrix \matr\ is invertible, and it is easy to show that 
the inverse is given by the last line of \matr\ but after 
substituting $P_{{\vec k}}(q)$ by 
$1/(P_{{\vec k}}(q))$. This means that, once we have 
computed all the $f_{R}$ for a fixed 
$\ell$, we can obtain the new generating functions 
${\widehat f}_{R}(q, \lambda)$ using the explicit 
expression for $M^{-1}_{R\, R'}(q)$. 
Clearly, from the Chern-Simons point of view it is highly nontrivial 
that the polynomials obtained in this way have the structure \tildefr.   
   
So far we have analyzed the representation and spin content
arising from a D2 brane whose moduli can change
in a Calabi-Yau manifold, but we have assumed
that its topology (the genus $g$ and the number of holes)
does not change. This is not the most general situation. However
as was shown in the closed string case \gvm\ , by
a physical reasoning in target space, studying D2 branes
as if they have a fixed genus gives a structure for the topological
string answer that is in fact the generic case.  This has also
been understood more directly by studying the degeneration
structure of Riemann surfaces in the moduli space of D2 brane
\kkv .  
  It is plausible to assume that the same holds true
 for the open string case and that the expression we have found for 
$f_R(q, \lambda )$ in terms of integers ${\widehat N}_{R,g,Q}$
is more generally valid.  It would be interesting to verify this.
Note, for example, that when we write $g$ 
in ${\widehat N}_{R,g,Q}$ we do
not strictly mean a genus $g$ surface.  More precisely
this can arise from a D2 brane with genus bigger than or equal
to $g$, which at some points along the moduli degenerates 
to a genus $g$ surface, as in \kkv .  Also the number of 
holes on the Riemann surface is greater than or equal to the number of
boxes
in $R$.

On the physics side we will have to develop what these integers
directly compute.  In particular the notion of the ``bulk spin''
$g$ must have an intrinsic physical meaning in the target
space description.  
  Moreover one should be able to understand
directly from target space reasoning how the ${\bf S}_R (H^{\ell-1})$
arises
and how the Clebsch-Gordon coefficients $C_{RR'R''}$ arise.  At any
rate we will find further evidence, from the computations of
link invariants of Chern-Simons theory that this picture
is correct.

One important advantage of the formulation \fr\
 is that the relation \nrel\ is now manifest. From \coefchar, 
it follows that $C_{R \, R'^t \, R''^t}= C_{R\,R'\,R''}$, and using 
\lrschur, one easily derives \nrel. Notice that, in contrast 
to the integers $N_{R,Q,s}$, the new integers 
${\widehat N}_{R,g,Q}$ are all independent. This 
shows that, although the generating functions 
\tildefr\ and \fr\ are equivalent (since they are related by 
an invertible matrix), 
 the invariants introduced in \ints\ are more fundamental 
than the integers appearing in \ovconj.

Before exploring further consequences of the structure result \fr, 
let's write it in some detail for representations with $\ell=1,2,3$. 
For $\ell=1$, we have:
\eqn\fone{
f_{\tableau{1}}(q, \lambda)=\sum_Q \sum_{g\ge 0}  {\widehat N}_{R,g,Q}
(q^{-{1\over 2}} -q^{1 \over 2})^{2g-1} \lambda^Q,}
where we have taken into account that $S_{\tableau{1}}(q)=1$. Since 
$f_{\tableau{1}}(\lambda,q)$ is just the unnormalized HOMFLY polynomial 
of the knot \lm, the above result is in perfect 
agreement with the fact that the HOMFLY polynomial can be 
written in terms of even powers of the variable 
 $q^{-{1\over 2}} -q^{1 \over 2}$ \lick\lickm. For $\ell=2$, we have 
\eqn\eletwo{
\eqalign{
f_{\tableau{2}}(q, \lambda)=&\sum_Q \sum_{g\ge 0} 
\bigl( q^{-{1 \over 2}} {\widehat N}_{\tableau{2},g,Q} - 
 q^{1 \over 2} {\widehat N}_{\tableau{1 1},g,Q}\bigr)
(q^{-{1\over 2}} -q^{1 \over 2})^{2g-1} \lambda^Q,\cr 
f_{\tableau{1 1}}(q, \lambda)=&\sum_Q \sum_{g\ge 0} 
\bigl( -q^{{1 \over 2}} {\widehat N}_{\tableau{2},g,Q} + 
 q^{-{1 \over 2}} {\widehat N}_{\tableau{1 1},g,Q}\bigr)
(q^{-{1\over 2}} -q^{1 \over 2})^{2g-1} \lambda^Q.\cr}}
Finally, for $\ell=3$, we find:
\eqn\elethree{
\eqalign{
f_{\tableau{3}}(q,\lambda)=&\sum_Q \sum_{g \ge 0} 
\bigl( q^{-1} {\widehat N}_{\tableau{3},g,Q} - 
 {\widehat N}_{\tableau{2 1},g,Q} + 
q {\widehat N}_{\tableau{1 1 1},g,Q} \bigr)
(q^{-{1\over 2}} -q^{1 \over 2})^{2g-1} \lambda^Q,\cr 
f_{\tableau{2 1}}(q, \lambda)=&\sum_Q \sum_{g \ge 0} 
\bigl(-{\widehat N}_{\tableau{3},g,Q}  
+ (q+q^{-1}-1){\widehat N}_{\tableau{2 1},g,Q}-
{\widehat N}_{\tableau{1 1 1},g,Q}\bigr)
(q^{-{1\over 2}} -q^{1 \over 2})^{2g-1} \lambda^Q,\cr 
f_{\tableau{1 1 1}}(q, \lambda)=&\sum_Q \sum_{g \ge 0} 
\bigl( q {\widehat N}_{\tableau{3},g,Q} - 
 {\widehat N}_{\tableau{2 1},g,Q} + 
q^{-1} {\widehat N}_{\tableau{1 1 1},g,Q} \bigr)
(q^{-{1\over 2}} -q^{1 \over 2})^{2g-1} \lambda^Q.\cr}}  

One important consequence of \fr\ is 
that it encodes the structure of the large $N$ expansion \frek.
To show this, it is very convenient 
to transform the $f_{R}$ functions to the $\vec k$-basis. 
We then define: 
\eqn\rele{
f_{\vec k}(q, \lambda)
=\sum_{R} 
\chi_{R}(C(\vec k)) f_{R}(q,\lambda).}
Notice that, in terms of these functions, the equation 
\master\ reads
\eqn\stfk{G^{(c)}_{\vec k}(q,\lambda) =
\sum_{n|\vec k} 
n^{|\vec k|-1} 
f_{\vec k_{1/n}}(q^n, \lambda^n).}
This implies that the functions 
$f_{\vec k}(q, \lambda)$ are 
given by the connected Green functions, corrected by some lower order 
terms that involve $f_{\vec k'}$ with $\ell' \le \ell$. Since 
\eqn\weakpre{
{|C(\vec k)|\over \ell!} \, G_{\vec k}^{(c)}(U)= 
\sum_{g=0}^\infty F_{g,\vec k}(\lambda)\, g_s^{2g-2+|\vec k|},}
to prove that \fr\ explains the relations between the $N_{R,Q,s}$ 
observed in \lm, one has to show that the right hand side of \stfk\ 
has an expansion in $g_s$ of the form \weakpre.
   
The proof goes as follows. Define the following integers:
\eqn\kints{
n_{{\vec k},g,Q}= 
\sum_{R} 
\chi_{R} (C({\vec k})) 
{\widehat N}_{R,g,Q}.}
Using again the expression \coefchar, and \pk\expk, it 
is easy to prove    
that the $f_{\vec k}(q, \lambda)$ have 
the following structure:
\eqn\stf{
f_{\vec k}(q, \lambda)=\biggl( {
\prod_j (q^{-{j\over 2}} -q^{{j\over 2}})^{k_j}  
\over (q^{-{1\over 2}}-q^{1\over 2})^2}\biggr)
\sum_{Q}\sum_{g\ge 0} n_{{\vec k},g,Q}
(q^{-{1\over 2}}-q^{{1\over 2}})^{2g}\lambda^Q.} 
We can now show that this
target space prediction gives the $1/N$ expansion \frek\ for 
the connected vevs. Notice first that 
\eqn\ttone{
f_{\vec k}(q^{-1}, \lambda)= 
(-1)^{
|k|} f_{\vec k}(q, \lambda).}
If we expand in $ g_s$, the 
leading power comes from the overall fraction in \stf: 
\eqn\exps{ {
\prod_j (q^{j\over 2} -q^{-{j\over 2}})^{ k_j}  
\over (q^{1\over 2}-q^{-{1\over 2}})^2}= 
g_s^{|\vec k|-2} 
+ \cdots. }
Since $f_{\vec k}(q, \lambda)$ has 
parity $(-1)^{|\vec k|}$ under $g_s\leftrightarrow -g_s$, 
 the powers of $g_s$ in the expansion are 
$2n+ |\vec k|-2$, with $n\ge 0$. 
This is precisely the structure of \frek. 
Using the definition of $\vec k_{1/n}$, it is 
easy to see that the terms $f_{\vec k_{1/n}}(q^n, \lambda^n)$ 
in \stfk, for $n>1$, have the 
same perturbative expansion in $g_s$, 
and this finally shows that the structure of the 
expansion \frek\ is a consequence of \stf. We conclude 
that the structure theorem \fr\ encodes all the relations between 
the integers $N_{R,Q,s}$ found in \lm.    

As a final remark, it is interesting to give the geometric interpretation 
of the integers defined in \kints. If the symmetric group $S_{\ell}$ is 
acting on a manifold $M$, the fixed points of the action will be 
labeled by the conjugacy classes $C(\vec k)$. We will denote them by 
$M^{\vec k}$. One can then consider the invariant part of the cohomology 
of $M$ under the Young symmetrizer $c_R$. The fixed point theorem 
 then tells us that
\eqn\fpoint{
\chi({\bf S}_R(H^*(M)))=\sum_{\vec k} {|C(\vec k)| \over \ell!} 
\chi_R(C(\vec k)) \chi( M^{\vec k}).} 
This is just the expression of the fact that if 
the group $G$ acts on $M$, the number of cohomology elements of
$M$ in the representation $R$ of $G$ are given by the
sum of cohomologies of subspaces of $M$ fixed by each element
of $h\in G$, weighted by the character of $h$ in representation $R$,
$\chi_R(h)/|G|$.
In our context, $M={\cal M}_{g,\ell}$, and the symmetry group acts 
by permutation of the boundaries. The integer 
$n_{{\vec k},g,Q}$ is then interpreted as the 
Euler characteristic of the fixed point subset of ${\cal M}_{g,\ell}$ 
labeled by $\vec k =(k_1, k_2, \cdots)$. In this subset we have 
$k_1$ simple boundaries, $k_2$ double boundaries, and in general 
$k_j$ boundaries made up of $j$ boundaries that have collided. Therefore, 
this geometric configuration has effectively $h=\sum_j k_j$ boundaries, 
as in the $1/N$ expansion \frek.

\newsec{Links and topological strings}

\subsec{Generalization to links}

In this section, we will generalize the construction of \ov\lm\ and of the 
above sections to links.  
Let's consider a link ${\cal L}$ with $L$ components $\gamma_{\alpha}$, 
$\alpha=1, \cdots, L$. As explained in section 2, the 
BPS states will be classified now 
by the quantum numbers $Q$, the spin $s$ and the representations 
$R_1, \cdots, R_L$ associated to the D2 branes ending on the 
Lagrangian submanifolds
$D_{\alpha}$. The integers associated to 
these BPS states will be denoted by $N_{(R_1, \cdots, R_L);Q,s}$. The free 
energy is then given by 
\eqn\free{
F(V_{\alpha})=\sum_{n=1}^{\infty}\sum_{R_1, \cdots, R_L} \sum_{Q,s} 
{N_{(R_1, \cdots, R_L);Q,s}\over q^{n\over2}-q^{-{n\over2}}}{q^{ns} 
\lambda^{nQ} \over n}
\prod_{\alpha=1}^L {\rm Tr}_{R_\alpha}V_{\alpha}^n.}
The main conjecture is then that 
\eqn\conj{
\langle Z(U_1, \cdots, U_L; V_1, \cdots, V_L) \rangle = 
\langle \exp\bigl[
\sum_{\alpha=1}^L \sum_{n=1}^\infty {1 \over n} {\rm Tr}\, U_{\alpha}^n\, 
{\rm Tr }\, V_{\alpha}^n\bigr]\rangle =
\exp F( V_{\alpha}).}

The generalization of \master\ is straightforward. The basis of 
operators is labeled now by vectors $\vec k^{(\alpha)}$, 
$\alpha=1, \cdots, L$, and we define the vev 
$G_{\vec k^{(1)}, \cdots,\vec k^{(L)} }(U_1, 
\cdots, U_L)$ as follows:
\eqn\linkvev{
G_{\vec k^{(1)}, \cdots,\vec k^{(L)} }(U_1, 
\cdots, U_L)=\langle \prod_{\alpha=1}^L\Upsilon_{\vec k^{(\alpha)}}
(U_{\alpha}) \rangle, }
where $U_{\alpha}$ is the holonomy around ${\cal K}_{\alpha}$.  
Using Frobenius formula, we can obtain the representation 
basis for the Wilson loop operators:
\eqn\froblink{
 {\rm Tr}_{R_1}(U_1) \cdots {\rm Tr}_{R_L}(U_L)= \prod_{\alpha=1}^L 
 \chi_{R_{\alpha}}(C(\vec k^{(\alpha)}))\Upsilon_{\vec k^{(\alpha)}}
(U_{\alpha}). }
The connected vevs are defined, as usual, by taking the 
logarithm of the generating function. More precisely, 
associate to any $\vec k$ the polynomial $p_{\vec k}(x)= \prod_j
x_j^{k_j}$ 
in the variables $x_1, x_2, \dots$. We then define the 
``connected" coefficients $a_{\vec k^{(1)}, \cdots,\vec k^{(L)} }$ 
as follows:  
\eqn\cumulink{
\eqalign{
&\log\Bigg(1+ \sum_L \sum_{\vec k^{(\alpha)}} a_{\vec k^{(1)}, 
\cdots,\vec k^{(L)} }
\prod_{\alpha=1}^L {|C(\vec k^{(\alpha)})| \over \ell_{\alpha}!}p_{\vec k
^{(\alpha)}}(x^{(\alpha)})
 \Bigg)=\cr 
& \,\,\,\,\,\,\,\,\,\ 
\sum_L\sum_{\vec k^{(\alpha)}} a^{(c)}_{\vec k^{(1)}, \cdots,\vec k^{(L)}
}
\prod_{\alpha=1}^L
{|C(\vec k^{(\alpha)})| \over \ell!}p_{\vec k^{(\alpha)}}(x^{(\alpha)}).
\cr}}  
It is again convenient to group the numbers
 $N_{(R_1, \cdots, R_L);Q,s}$ into the polynomials
\eqn\polyn{
f_{(R_1, R_2, \cdots, R_L)}(q, \lambda)= \sum_{Q,s} 
{N_{(R_1, \cdots, R_L);Q,s}\over q^{1\over2}-q^{-{1\over2}}}q^s \lambda^Q
.}
A simple generalization of the arguments of \lm\ gives the following 
relation for links:
\eqn\masterlinks{
G_{\vec k^{(1)}, \cdots,\vec k^{(L)} }^{(c)}(U_1, 
\cdots, U_L)=\sum_{n|\vec k^{(\alpha)}} 
n^{\sum_{\alpha} |\vec k^{(\alpha)}|-1} \sum_{R_1, \cdots, R_L} 
\prod_{\alpha=1}^L \chi_{R_{\alpha}}(C(\vec k_{1/n}^{(\alpha)})) 
f_{(R_1, \cdots, R_L)}(q, \lambda).} 
In the first sum, the vector $\vec k_{1/n}$ is defined as in \master. 
In \masterlinks, the $n$ have to divide 
all the vectors $\vec k^{(\alpha)}$, $\alpha=1, \cdots, L$.  
As in \lm, it is easy to see that, given 
the connected vevs, the equation \masterlinks\ determines the 
generating functions  
$f_{(R_1, \cdots, R_L)}(q, \lambda)$ in a unique way by a 
recursive procedure. For every vector $(\ell_1, \cdots, \ell_L)$ 
(which specifies the ``order'') we can solve for the $f$'s in terms 
of the connected vevs and the $f$'s of lower order ({\it i.e.} with 
$\ell_{\alpha}'\le\ell_{\alpha}$).

Let us now give some examples of this procedure.
The simplest case of \masterlinks\ is for $R_1 = 
\cdots =R_L= \tableau{1}$. In this case, one also has 
$\vec k^{(1)} =\cdots \vec k^{(L)}=(1,0,\cdots)$, and 
\eqn\simplemas{
f_{(\tableau{1}, \cdots, \tableau{1})}=G^{(c)}_{(1,0,\cdots), 
\cdots, (1,0,\cdots)}(U_1, \cdots, U_L).}
Notice that the connected vacuum expectation value contains 
information also about the vevs for the components of the link, 
and for all the possible sublinks that can be formed with these 
components. For example, for links with two components, one has
\eqn\twoc{
f_{(\tableau{1}, \tableau{1})}=\langle {\rm Tr}\, U_1 \, 
{\rm Tr}\, U_2 \rangle -\langle {\rm Tr} U_1 \rangle \langle 
{\rm Tr} U_2 \rangle,}
where the second piece is the right hand side is the product 
of the vevs for the two knots, ${\cal K}_1$ and ${\cal K}_2$, 
that form the link. 

Let us now focus on links with two components. The next order is specified 
by $(\ell_1, \ell_2)=(2,1)$ and $(\ell_1, \ell_2)=(1,2)$. We have, for 
example, 
\eqn\ordertwo{
\eqalign{
f_{(\tableau{2},\tableau{1})}=&{1 \over 2}\bigl( G^{(c)}_{(2,0,\cdots), 
(1,0,\cdots)}+G^{(c)}_{(0,1,0,\cdots), 
(1,0,\cdots)}\bigr),\cr
f_{(\tableau{1 1},\tableau{1})}=&{1 \over 2}\bigl( G^{(c)}_{(2,0,\cdots), 
(1,0,\cdots)}-G^{(c)}_{(0,1,0,\cdots), 
(1,0,\cdots)}\bigr).\cr}}
Of course, for $(\ell_1, \ell_2)=(1,2)$ we have the same equations 
but with the labels exchanged. 

The next level is labeled by $(1,3)$, $(3,1)$ and $(2,2)$. For $(3,1)$, 
\masterlinks\ gives:
\eqn\levelthreeone{
\eqalign{
f_{(\tableau{3},\tableau{1})}=&{1\over
6}G^{(c)}_{(3,0,\cdots),(1,0,\cdots)}
+{1\over 2}G^{(c)}_{(1,1,0,\cdots),(1,0,\cdots)} +
 {1\over 3}G^{(c)}_{(0,0,1,\cdots),(1,0,\cdots)},\cr
f_{(\tableau{2 1},\tableau{1})}=&{1\over 3}\bigl( 
G^{(c)}_{(3,0,\cdots),(1,0,\cdots)}
-G^{(c)}_{(0,0,1,\cdots),(1,0,\cdots)}\bigr),\cr
f_{(\tableau{1 1 1},\tableau{1})}=&
{1\over 6}G^{(c)}_{(3,0,\cdots),(1,0,\cdots)}
-{1\over 2}G^{(c)}_{(1,1,0,\cdots),(1,0,\cdots)} 
+{1\over 3}G^{(c)}_{(0,0,1,\cdots),(1,0,\cdots)}.\cr}}
The equations for the order $(1,2)$ are again obtained by 
exchanging the labels. Finally, for $(2,2)$ we obtain 
 \eqn\leveltwotwo{
\eqalign{
f_{(\tableau{2},\tableau{2})}(q,\lambda)=&{1\over 4}\bigl( 
G^{(c)}_{(2,0,\cdots),(2,0,\cdots)}
+G^{(c)}_{(0,1,0,\cdots),(2,0,\cdots)} +
 G^{(c)}_{(2,0,\cdots),(0,1,\cdots)}+
G^{(c)}_{(0,1,\cdots),(0,1,\cdots)}\bigr)
\cr & -
{1 \over 2}f_{(\tableau{1},\tableau{1})}(q^2,\lambda^2) ,\cr
f_{(\tableau{2},\tableau{1 1})}(q, \lambda)=&{1\over 4}
\bigl( 
G^{(c)}_{(2,0,\cdots),(2,0,\cdots)}
+G^{(c)}_{(0,1,0,\cdots),(2,0,\cdots)} -
 G^{(c)}_{(2,0,\cdots),(0,1,\cdots)}-
 G^{(c)}_{(0,1,\cdots),(0,1,\cdots)}\bigr) \cr & +
{1 \over 2}f_{(\tableau{1},\tableau{1})}(q^2,\lambda^2)\cr
f_{(\tableau{1 1},\tableau{2})}(q, \lambda)=&{1\over 4}
\bigl( 
G^{(c)}_{(2,0,\cdots),(2,0,\cdots)}-
G^{(c)}_{(0,1,0,\cdots),(2,0,\cdots)} +
 G^{(c)}_{(2,0,\cdots),(0,1,\cdots)}-
 G^{(c)}_{(0,1,\cdots),(0,1,\cdots)}\bigr) \cr & +
{1 \over 2}f_{(\tableau{1},\tableau{1})}(t^2,\lambda^2), 
\cr 
f_{(\tableau{1 1},\tableau{1 1})}(q, \lambda)=&{1\over 4}
\bigl( 
G^{(c)}_{(2,0,\cdots),(2,0,\cdots)}-
G^{(c)}_{(0,1,0,\cdots),(2,0,\cdots)} -
 G^{(c)}_{(2,0,\cdots),(0,1,\cdots)}+
 G^{(c)}_{(0,1,\cdots),(0,1,\cdots)}\bigr)\cr & -
{1 \over 2}f_{(\tableau{1},\tableau{1})}(q^2,\lambda^2).
\cr}} 

\subsec{D-branes and links}

A straightforward generalization of the arguments in section 
2 goes as follows. If we have a link of $L$ 
components, the corresponding Lagrangian submanifold $D=\cup D_\alpha $ 
after the transition
will have $b_1=L$. We will denote by $Y_{\alpha}$ 
the nontrivial one-cycles of $D$, 
with $\alpha=1, \cdots, L$. After performing the 
analytic continuation, the $1/N$ expansion \mini\ can 
be written as
\eqn\largentwolinks{
F(V_{\alpha})=\sum_{{\vec k}^{(1)}, \cdots,{\vec k}^{(L)}} 
\sum_{g=0}^{\infty} g_s^{2g-2+\sum_{\alpha=1}^L 
|\vec k^{(\alpha)}|} F_{g;  \vec k^{(1)}, \cdots,\vec k^{(L)} }
(\lambda) \Upsilon_{\vec k^{(\alpha)}}
(V_{\alpha})
,}
where $\vec k^\alpha$ denotes the number
of holes, together with their wrappings which end on $D_{\alpha}$. 

The D-brane derivation of this expansion proceeds as in section 3. 
We have 
to consider Riemann surfaces with 
$\ell=\sum_{\alpha=1}^L \ell_{\alpha}$ boundaries in such a way that 
$\ell_{\alpha}$ boundaries end on $Y_{\alpha}$. The relevant 
symmetry group is now $S_{\ell_1} \times \cdots \times S_{\ell_L}$, and 
the starting point is again \conf. Projecting onto 
symmetric configurations under the action of 
$\prod_{\alpha=1}^L S_{\ell_{\alpha}}$, one finds:
\eqn\decom{
\sum_{R_1, \cdots, R_L} \biggl( {\bf S}_{R_1}(F^{\ell_1}) \otimes 
\cdots \otimes {\bf S}_{R_L}(F^{\ell_L}) \biggr) \otimes 
{\bf S}_{R_1, \cdots, R_L}(H^*(({\bf S}^1)^{\ell-1}) \otimes 
H^*({\cal M}_{g,\ell})).} 
The first factor is simply the tensor product of the 
irreducible representations $R_1, \cdots, R_L$ of $SU(N)$, while the
second 
factor can be further decomposed as
\eqn\furt{
\sum_{R'_1, R_1'' \cdots, R'_L, R_L''} C_{R_1\, R'_1\, R''_1} 
\cdots C_{R_L\, R'_L\, R''_L} \, 
{\bf S}_{R'_1, \cdots, R'_L}(H^*(({\bf S}^1)^{\ell-1})) \otimes 
{\bf S}_{R''_1, \cdots, R''_L}(H^*({\cal M}_{g,\ell})) 
}
The generating functions associated to 
${\bf S}_{R'_1, \cdots, R'_L}(H^*(({\bf S}^1)^{\ell-1}))$ 
are given by a straightforward generalization of the 
arguments for the $L=1$ case. The result is:
\eqn\genlink{
{\rm Tr}_{{\bf S}_{R_1, \cdots, R_L}(H^*(({\bf S}^1)^{\ell-1}))} 
(-1)^F q^s =
(q^{-{1\over 2}}-q^{1 \over 2})^{L-1} \prod_{\alpha=1}^L
S_{R_{\alpha}}(q).}
The new integer invariants will then be, in the case of links,  
\eqn\intslinks{
{\widehat N}_{(R_1, \cdots, R_L),g,Q}=\chi 
\bigl( {\bf S}_{R_1, \cdots, R_L}(H^*({\cal M}_{g,\ell}))\bigr),}
and the generating functions \polyn\ have the structure:
\eqn\frlinks{
\eqalign{
&f_{(R_1, \cdots, R_L)}(q, \lambda)=\cr & 
(q^{-{1\over 2}}-q^{1 \over 2})^{L-2}
\sum_{g\ge 0} \sum_{Q}
\sum_{R'_1, R_1'' \cdots, R'_L, R_L''}  \prod_{\alpha=1}^L
C_{R_{\alpha}\,R_{\alpha}'\,R_{\alpha}''}S_{R_{\alpha}'}(q) 
{\widehat N}_{(R_1'', \cdots, R_L''),g,Q} 
 (q^{-{1\over 2}}-q^{1\over 2})^{2g}\lambda^Q.\cr}}
We can also define
\eqn\tildefrlinks{
{\widehat f}_{(R_1, \cdots, R_L)}(q, \lambda)= 
(q^{-{1\over 2}}-q^{1 \over 2})^{L-2}
\sum_{g \ge 0}\sum_Q {\widehat N}_{(R_1, \cdots, R_L),g,Q}
(q^{-{1\over 2}}-q^{1 \over 2})^{2g}\lambda^Q.} We then have 
the relation:
\eqn\relafslinks{ 
 f_{(R_1, \cdots, R_L)}(q, \lambda)=\sum_{R_1', \cdots, R'_L} 
M_{R_1, \cdots, R_L;R'_1, \cdots, R_L'}(q) 
{\widehat f}_{(R_1, \cdots, R_L)}(q, \lambda),}
where the matrix $M_{R_1, \cdots, R_L;R'_1, \cdots, R_L'}(q)$ is given by
\eqn\matrlinks{
\eqalign{
M_{R_1, \cdots, R_L;R'_1, \cdots, R_L'}(q)&=
\sum_{ R_1'', \cdots, R_L''}  \prod_{\alpha=1}^L
C_{R_{\alpha}\,R_{\alpha}'\,R_{\alpha}''}S_{R_{\alpha}''}(q) \cr
 &=
\sum_{{\vec k}^{(1)}, \cdots, {\vec k}^{(L)}} 
\prod_{\alpha=1}^L {|C({\vec k}^{(\alpha)})| \over \ell_{\alpha}!} 
\chi_{R_{\alpha}}(C({\vec k}^{(\alpha)})) \chi_{R'_{\alpha}}
(C({\vec k}^{(\alpha)}))P_{{\vec k}^{(\alpha)}}(q).\cr}}
The generalization of \nrel\ to the case of links is
\eqn\numberrl{
N_{(R_1, \cdots, R_L);Q,s}=(-1)^{\sum_{\alpha=1}^L \ell_{\alpha} -1} 
N_{(R^t_1, \cdots, R^t_L);Q,-s},} 
or, equivalently, 
\eqn\rlefs{
f_{(R_1, \cdots, R_L)}(t^{-1},\lambda)=(-1)^{\sum_{\alpha=1}^L
\ell_{\alpha}}
f_{(R^t_1, \cdots, R^t_L)}(t,\lambda), }
which are a consequence of \lrschur\ and \frlinks. 
Again, we can obtain the structure of the 
generating functions in the $\vec k$ basis, by defining
\eqn\relelinks{
f_{\vec k^{(1)}, \cdots, \vec k^{(L)}}(q, \lambda)
=\sum_{R_1, \cdots, R_L} 
\prod_{\alpha=1}^L 
\chi_{R_{\alpha}}(C(\vec k^{(\alpha)})) f_{(R_1, \cdots,
R_L)}(q,\lambda).}
and
\eqn\kintslinks{
n_{({\vec k}^{(1)}, \cdots, {\vec k}^{(L)}),g,Q}= 
\sum_{R_1, \cdots, R_L} \prod_{\alpha=1}^L
\chi_{R_{\alpha}} (C({\vec k}^{(\alpha)})) 
{\widehat N}_{(R_1, \cdots, R_L),g,Q}.} 
A straightforward generalization of the argument in the 
case of knots shows that 
\eqn\stflinks{
f_{\vec k^{(1)}, \cdots, \vec k^{(L)}}(q, \lambda)=\biggl( {
\prod_j (q^{-{j\over 2}} -q^{{j\over 2}})^{\sum_{\alpha=1}^L
k_j^{(\alpha)}}  
\over (q^{-{1\over 2}}-q^{1\over 2})^2}\biggr)
\sum_{Q}\sum_{g\ge 0} n_{(\vec k^{(1)}, \cdots, \vec k^{(L)}),g,Q}
(q^{-{1\over 2}}-q^{{1\over 2}})^{2g}\lambda^Q.} 
and this leads to the $1/N$ expansion \largentwolinks. 

\subsec{Links in Chern-Simons theory}

In the previous section we have generalized the results of \ov\lm\ and 
section 3 to the case of links in a rather straightforward way. 
The subtlety with links has to do rather with the computation of the vevs 
in Chern-Simons theory. It turns out that there are some 
ambiguities involved in this computation, and as we will see the 
appropriate definition of the vev is different from the standard one 
in knot theory. 

A good starting point to address this issue 
is the perturbative analysis of vevs in Chern-Simons 
theory \gmm\alp. Let us consider a link ${\cal L}$ with $L$ components 
${\cal K}_{\alpha}$, $\alpha=1, \cdots, L$, and let's assume for
simplicity 
that all the components carry the 
fundamental representation. In Chern-Simons theory the 
components of a link have to be 
framed and the resulting invariants are invariants of framed links. 
We will assume that the components are in the vertical 
framing, which is very natural when we define the invariant in terms 
of a plane projection of the link \ilm. Vevs computed in the 
vertical framing are not ambient 
isotopy invariants, but rather regular isotopy invariant (this means 
that they are not invariant under Reidemeister type I moves). 
However, a perturbative analysis of the vev \gmm\alp\ 
shows that the corrected quantity  
\eqn\corquan{
\exp \bigl[ -2 \pi i w({\cal L})h_{\tableau{1}} \bigr]\langle W({\cal L}) 
\rangle^{\rm vf},}
where $w({\cal L})$ is the writhe of the planar projection of the link, is 
an ambient isotopy invariant and in particular gives the unnormalized 
HOMFLY polynomial of the link. In this equation, 
$h_R$ is the conformal weight of the WZW model for the representation 
$R$. For the fundamental representation it is given by 
$h_{\tableau{1}}=(N-1/N)/(2(k+N))$. The key thing here 
(which is familiar from 
the construction of the Kauffman bracket) is that the variation of 
$\langle W({\cal L}) \rangle^{\rm vf}$ under type I moves is compensated 
by the variation of the writhe, in such a way that the product 
is a topological invariant. Notice that the writhe of 
the link is given 
by:
\eqn\writhedec{
w({\cal L})= \sum_{\alpha=1}^L w({\cal K}_{\alpha}) + 
2\,  {\rm lk}({\cal L}),}
where ${\rm lk}({\cal L})$, the total linking number of ${\cal L}$, is 
defined as:
\eqn\totalink{
{\rm lk}({\cal L})=
\sum_{\alpha <\beta} {\rm lk}({\cal K}_{\alpha}, {\cal K}_{\beta}),}
and ${\rm lk}({\cal K}_{\alpha}, {\cal K}_{\beta})$ 
is the linking number of the two
components $\alpha$, $\beta$. These linking numbers are 
ambient isotopy invariants of the link.
Therefore, in order to 
obtain a well-defined invariant of links, it would have been enough 
to introduce the correction factor $\exp \bigl[ -2 \pi i 
\sum_{\alpha}w({\cal K}_{\alpha})h_{\tableau{1}} \bigr]$. From a
topological 
point of view, there is an important advantage in defining 
the invariant as in \corquan: with that definition, 
all the crossings in the link are treated in the same way, without 
taking into account if they belong to the same component or to 
different components of the link (for a related discussion, 
see \ilr). This is why the usual HOMFLY polynomial 
satisfies a universal skein relation. 

The main conclusion of this analysis is that, in order to obtain the 
usual knot invariants, one has to correct the field theory vev 
in two ways. First, as for knots, we have to include a non-topological 
factor to take into account the framing of the components. The resulting 
vev is in the standard framing and will be denoted by 
$\langle W({\cal L})\rangle_{\rm sf}$:
\eqn\sframe{
\langle W({\cal L})\rangle^{\rm sf}=\exp \bigl[ -2 \pi i 
\sum_{\alpha}w({\cal K}_{\alpha})h_{\tableau{1}} \bigr]
\langle W({\cal L})\rangle^{\rm vf}.}
The second factor is topological and depends on the 
linking numbers of the components:  
\eqn\factor{
\exp\bigr[-4\pi i h_{\tableau{1}}\sum_{\alpha<\beta}{\rm lk}
({\cal K}_{\alpha}, {\cal K}_{\beta})\bigl].} 
Notice that \factor\ can be written as:
\eqn\rewr{
\exp\biggr[ {2 \pi i \over N(k+N)} 
\sum_{\alpha<\beta}{\rm lk}({\cal K}_{\alpha}, {\cal K}_{\beta})\biggl]
 \lambda^{-\sum_{\alpha<\beta}{\rm lk}({\cal K}_{\alpha}, 
{\cal K}_{\beta})},}
and the first factor guarantees that the corrected vev is a 
polynomial in $t$ and $\lambda$. 
However, this correction has some unpleasant features. It can be easily 
seen that the second factor spoils the structure of the
 $1/N$ expansion \frek\  
of the field-theory vev. This seems 
to be a problem in order to find a string interpretation of the 
Chern-Simons vevs. On the other hand, the string interpretation makes 
clear that the appropriate vevs should be functions of $t$ and $\lambda$,
so 
the vev  still has to be corrected with the first factor of 
\rewr\ in order to compare it 
to the string predictions. Therefore, we propose that the relevant 
vev that is needed for comparison with the string predictions is
\eqn\rightfact{
\exp\biggr[ {2 \pi i \over N(k+N)}\sum_{\alpha<\beta}{\rm lk}
({\cal K}_{\alpha}, {\cal K}_{\beta})\biggl]
\langle W({\cal L})\rangle^{\rm sf}.}
This means, in 
particular, that the relevant link polynomial is {\it not} the 
HOMFLY polynomial 
$P_{\cal L}(t, \lambda)$, but rather
\eqn\modhom{
\lambda^{{\rm lk}({\cal L})}P_{\cal L}(t, \lambda).} 
As we will see, this is crucial in order to find agreement with the 
string predictions. 

In the discussion above, we have assumed that all the components 
of the links were in the fundamental representation. We will now consider 
the general case in which the components are in arbitrary irreducible 
representations of $SU(N)$. To do that, we will introduce some notation. 
Let $R$ be an irreducible representation, and let $\Lambda$ be the highest 
weight. If we denote by 
$\lambda_i$, $i=1, \cdots, N-1$ the fundamental weights of $SU(N)$, we can 
always write 
\eqn\dec{
\Lambda=\sum_{i=1}^{N-1} a_i \lambda_i,}
where the $a_i$ are nonnegative integers. We can associate to $R$ a 
Young diagram in the usual way, with $\ell =\sum_i ia_i$ boxes. Consider
the 
following integer associated to $\Lambda$, which will be useful later on:  
\eqn\kap{
\kappa_{\Lambda}=\sum_i (ia_i^2 -i^2 a_i) + 2 \sum_{i<j}ia_ia_j.}
The conformal weight $h_R$ is then given by:
\eqn\cw{
h_R={\Lambda\cdot (\Lambda + 2 \rho) \over 2(k+N)},} 
where $\rho$ is the Weyl vector ({\it i.e.}, the sum of the 
fundamental weights). Notice that
\eqn\ident{
\Lambda\cdot (\Lambda + 2 \rho)=N\ell - {\ell^2\over N} +
\kappa_{\Lambda}.}
Let's then consider a ``multicoloured'' 
link with $L$ components, with  
${\cal K}_{\alpha}$ in the representation $R_{\alpha}$, for  
$\alpha=1, \cdots, L$. The number of boxes in the Young tableau 
associated to $R_{\alpha}$ is denoted by $\ell_{\alpha}$.   
The generalization of \sframe\ to this situation is simply
\eqn\sfrgen{
\langle W ({\cal L}) \rangle^{\rm sf}=\exp \bigl[ -2 \pi i 
\sum_{\alpha}w({\cal K}_{\alpha})h_{R_{\alpha}} \bigr]
\langle W({\cal L})\rangle^{\rm vf}.}  
We also propose that the 
corrected vev which extends \rightfact\ to the general case is 
\eqn\gencor{ 
\exp\biggr[ {2 \pi i \over N(k+N)}\sum_{\alpha<\beta}{\rm lk}
({\cal K}_{\alpha}, {\cal K}_{\beta})\ell_{\alpha}
\ell_{\beta}\biggl]\langle 
W({\cal L})\rangle^{\rm sf}.} 
This choice is motivated by the structure of the knot operators for 
torus knots and links, that will be reviewed later. \gencor\ is 
the minimal correction that makes the resulting object a function of  
$t^{\pm 1/2}$ and $\lambda^{\pm 1/2}$, and it can be easily 
checked that it preserves the 
structure of the $1/N$ expansion.   
 
\subsec{Rank-level duality for links} 

Rank-level duality of WZW models \wzw\ implies a certain number 
of identities for Chern-Simons vevs \ns, which essentially relate 
the vev of a Wilson loop in the representation $R$ with the vev 
in the transposed representation. In this section we will 
see that these identities are in fact closely related to the 
relation \rlefs. The comparison to 
the results of \ns\ involves the extra factor \gencor\ in an 
interesting way, and provides a further consistency check of our 
procedure.  

Using \rlefs\ and \masterlinks\ one can obtain the following relation for
the 
vevs of a product of Wilson loops in the representation basis:
\eqn\rllink{
\langle {\rm Tr}_{R_1}(U_1) \cdots {\rm Tr}_{R_L}(U_L)
\rangle (q^{-1},\lambda)= (-1)^{\ell} 
\langle {\rm Tr}_{R^t_1}(U_1) \cdots {\rm Tr}_{R^t_L}(U_L)
\rangle (q,\lambda),} 
where $\ell =\sum_{\alpha}\ell_{\alpha}$. 
This relation is in fact a consequence of the usual 
$1/N$ expansion of the Wilson loops in Chern-Simons theory, and 
of the fact that the vevs can be written in terms of the 
variables $q$, $\lambda$. Therefore, for this relation to 
be true it is crucial to introduce 
the correction factors \gencor. To make 
contact with the rank-level duality relations
 obtained in \ns, we have to go to the 
vertical framing. The vevs in 
the vertical framing are related to our corrected vevs 
through \sframe\gencor:
\eqn\relavf{
\eqalign{
&\langle {\rm Tr}_{R_1}(U_1) \cdots {\rm Tr}_{R_L}(U_L)
\rangle ^{\rm vf}=\cr
&\exp\biggl[-2 \pi i \sum_{\alpha}w({\cal K}_{\alpha}) 
h_{R_{\alpha}}-{2\pi i \over N(k+N)} \sum_{\alpha< \beta} 
{\rm lk}({\cal K}_{\alpha}, {\cal K}_{\beta})
\ell_{\alpha}\ell_{\beta}  \biggr] 
\langle {\rm Tr}_{R_1}(U_1) \cdots {\rm Tr}_{R_L}(U_L)
\rangle.\cr}} Now we use the fact that 
taking the mirror image ${\widetilde U}$ of a Wilson line 
$U$ is equivalent to complex 
conjugation and sends $q, \lambda \rightarrow q^{-1}, \lambda^{-1}$. 
Finally, using that \ns\
\eqn\confrl{
h_{R_{\alpha}}\big|_{SU(N)_k} + h_{R^t_{\alpha}}\big|_{SU(k)_N}= 
{\ell_{\alpha}\over 2} -{\ell_{\alpha}^2 \over 2Nk},}
we obtain 
\eqn\rllinks{
\eqalign{
& \langle {\rm Tr}_{R_1}(U_1) \cdots {\rm Tr}_{R_L}(U_L)
\rangle ^{\rm vf}_{SU(N)_k}=\cr
& \exp\biggl[ \pi i \sum_{\alpha}w({\cal K}_{\alpha}) 
\ell_{\alpha}-{2\pi i \over N(k+N)} \sum_{\alpha, \beta} 
{\rm lk}({\cal K}_{\alpha}, {\cal K}_{\beta})
 \ell_{\alpha}\ell_{\beta}  \biggr] 
\langle {\rm Tr}_{R^t_1}(\widetilde U_1) \cdots {\rm Tr}_{R^t_L}( 
\widetilde U_L)
\rangle^{\rm vf}_{SU(k)_N},\cr} }
where the self-linking number is ${\rm lk}({\cal K}_{\alpha}, 
{\cal K}_{\alpha})=w({\cal K}_{\alpha})$. To write the final result, 
we have also taken into account that the vev of 
${\rm Tr}_{R_1}(U_1) \cdots {\rm Tr}_{R_L}(U_L)$ has the monodromy 
$(-1)^{\ell}$ under $\lambda \rightarrow 
{\rm e}^{2\pi i}\lambda$. 
We then see that the rank-level duality 
relations of \ns\ are a consequence of both the structure of the 
$1/N$ expansion (or its D-brane version \stflinks) and 
the monodromy of the vevs under $\lambda \rightarrow 
{\rm e}^{2\pi i}\lambda$.

\subsec{Polynomial invariants for torus links}

To obtain explicit results for the generating functions 
$f_{(R_1, \cdots, R_L)}$, we have to 
compute vevs of general products of Wilson loops. In general 
these computations are technically difficult, but if the components 
of the links are torus knots, one can use the formalism of knot 
operators developed in \llr. Torus knots are labeled by two coprime 
integers $(n,m)$, which correspond to winding numbers around 
the two non-contractible cycles of the torus. Wilson loops corresponding
to 
a torus knot $(n,m)$, and in an irreducible representation 
of highest weight $\Lambda$, are represented by the operator 
$W_{\Lambda}^{(n,m)}$. This operator acts 
on the Hilbert space of Chern-Simons 
gauge theory on a torus. It has been shown in \quanta\ that 
this space has an orthonormal basis $|p\rangle$ labeled by 
weights $p$ in the fundamental chamber of the weight lattice 
of $SU(N)$, ${\cal F}_l$, where $l=k+N$. 
We take as representatives of $p$ the ones of the form 
$p=\sum_i p_i  \lambda_i$, 
with $p_i>0$ and $\sum_i p_i <l$. The vacuum is 
the state $| \rho \rangle$, where $\rho$ is the 
Weyl vector. The 
action of the loop operator on a state $|p \rangle$ is given by \llr: 
\eqn\knotop{
W_{\Lambda}^{(n,m)}|p\rangle = \sum_{\mu \in M_{\Lambda}}\exp \biggl[ 
-i\pi \mu ^2 {nm \over k+N} - 2\pi i {m \over k+N}  p \cdot  \mu 
\biggr] | p + n \mu\rangle.} 
If we act on the vacuum with $L$ knot operators $(n_1,m_1), \cdots,
(n_L,m_L)$, with representations labeled by 
$\Lambda_1, \cdots, \Lambda_L$, we will create a link whose $\alpha$-th 
component is a torus knot labeled by $(n_{\alpha}, m_{\alpha})$ 
and representation 
$\Lambda_{\alpha}$ \ilr. The resulting state 
can be computed from \knotop, and is 
given by
\eqn\act{
\eqalign{
& \biggl(\prod_{\alpha=1}^L W_{\Lambda_{\alpha}}^{(n_{\alpha},m_{\alpha})}
\biggr)| \rho \rangle = \sum_{\mu^{(\alpha)} \in M_{\Lambda_{\alpha}}}
\exp \biggl[ 
  - {2\pi i \over k+N}  
\bigl( \sum_{\alpha<\beta} n_{\alpha}m_{\beta}\, 
 \mu^{\alpha}\cdot \mu^{\beta} \bigr)\biggr]\cr 
&  \times\exp \biggl[-{i\pi \over k+N}\bigl 
(  \sum_{\alpha} n_{\alpha}m_{\alpha}
(\mu^{(\alpha)})^2 \bigr)-{2\pi i \over k+N}\rho\cdot 
\bigl(\sum_{\alpha}m_{\alpha}\mu^{(\alpha)}\bigr)
\biggr] | \rho + \sum_{\alpha}n_{\alpha} \mu^{(\alpha)} \rangle.\cr}} 
The link that has been constructed lives on the surface of a solid 
torus. To compute the vev of the product of Wilson loops when this 
link is in ${\bf S}^3$, we have to perform an $S$-transformation. The 
final expression for the vev, without including any correction factor, 
is \llr\ilr
\eqn\vevunnor{{\langle \rho | S \prod_{\alpha=1}^L 
W_{\Lambda_{\alpha}}^{(n_{\alpha},m_{\alpha})}
| \rho\rangle \over \langle  \rho | S| \rho\rangle}.} As explained in 
\lmtwo, the inner products that appear in \vevunnor\ can be computed as 
follows. The weight $\rho + \sum_{\alpha}n_{\alpha} \mu^{(\alpha)}$ 
is not necessarily in the fundamental chamber, but it will have a 
representative in it obtained by the action of an element of the 
Weyl group, say $w$. If this representative has a vanishing component 
in the Dynkin basis, then the corresponding state will be zero, due to 
the antisymmetry of the wave function under Weyl reflections. If this 
is not the case, we will be able to write the representative as 
$\rho + \mu_{({n_1}, \cdots, n_{\alpha})}$, where 
$\mu_{({n_1}, \cdots, n_{\alpha})}$ is a weight of nonnegative components. 
Using Weyl formula and the explicit expression for the $S$-operator, we
can 
write
\eqn\inner{
{\langle \rho | S 
| \rho + \sum_{\alpha}n_{\alpha} \mu^{(\alpha)}
\rangle \over \langle  \rho | S| \rho\rangle} 
=\epsilon(w){\rm ch}_{\mu_{({n_1}, \cdots, n_{\alpha})}}
\biggl[ -{2 \pi i \over 
k+N}\rho\biggr].}
This solves in principle the problem of computing the vev \vevunnor. 

We can now use these results to motivate \rightfact. 
First notice that the exponential in \act\ can not be written in terms of 
$t^{\pm 1/2}$, $\lambda^{\pm 1/2}$. This can be seen as follows. Let's 
denote by $\mu_i$, $i=1, \cdots, N$ the weights of the fundamental 
representation of $SU(N)$, and let $\Lambda$ be the highest weight of 
an irreducible representation of $SU(N)$ whose Young diagram 
contains $\ell$ boxes. It was shown in \lm\ that the weights in 
$M_{\Lambda}$ can be always written as  
\eqn\wpro{
k_1  \mu_{i_1} + \cdots + k_r \mu_{i_r}, \,\, 
1\le i_1< \dots <i_r\le N,}
where $(k_{\lambda})=(k_1, \dots, k_r)$ is an ordered 
partition of $\ell$, {\it i.e.} an $r$-tuple that sums up to $\ell$. 
Using this, and the explicit expression for 
the inner products $\mu_i \cdot \mu_j$ (see, for 
example, \lmtwo), it is easy to show that the second term in the 
exponential of \act\ gives a factor
\eqn\factorone{
\exp\biggl[ {\pi i \over N(k+N)} \sum_{\alpha}n_{\alpha}m_{\alpha} 
\ell_{\alpha}^2\biggr],}
while the first term gives
\eqn\factortwo{\exp\biggl[ {2 \pi i \over N(k+N)} \sum_{\alpha<\beta} 
n_{\alpha}m_{\beta} 
\ell_{\alpha}\ell_{\beta}\biggr].} 
None of these factors can be written in terms of $t^{\pm 1/2}$ 
and $\lambda^{\pm 1/2}$, since they involve a $1/N$ factor in the 
denominator of the exponent.  
As explained in \ilr, the writhe of each of the components of the link 
constructed in \act\ is $w({\cal K}_{\alpha})=n_{\alpha}m_{\alpha}$.  
Using \cw\ident\ it is easy to see that the factor \factorone\ will 
be canceled by the correction that is needed in order to enforce the 
standard framing. On the other hand, the remaining factor \factortwo\ 
is due to the linking of the different components, since 
${\rm lk}(K_{\alpha},K_{\beta})=-n_{\alpha}m_{\beta}$, $\alpha<\beta$, 
in this 
case \ilr. We see that the correction factor in \gencor\ is the minimal
one 
that has to be introduced in order to cancel \factortwo. Since the 
character involved in \inner\ is a rational function of $t^{\pm 1/2}$ 
and $\lambda^{\pm 1/2}$, it follows that the invariant defined in \gencor\ 
is also a rational function of these two variables, at least in the 
particular case of links made up of torus knots by the above construction. 

A particularly interesting example 
occurs when all the components of the link are the same torus knot. 
In this case, the resulting 
link is called a torus link. 
A torus link of $L$ components 
is made of $L$ torus knots of type $(n,m)$, and will be labeled 
by $(nL,mL)$, where $n$ and $m$ are relatively prime. 
For example, the Hopf link is the $(2,2)$ torus link. 
For torus links, the expression \act\ can be simplified very much. It is 
easy to see that when $(n_1,m_1) =\cdots =(n_L,m_L)=(n,m)$, the action of 
the string of operators on the vacuum is exactly that of a single 
knot operator $(n,m)$ but in the tensor product representation 
$\otimes_{\alpha=1}^L R_{\alpha}$. This means that the vev 
can be computed by decomposing the 
tensor product and using the results of \lm, without further ado. 
The decomposition in irreducible representations 
can be written as follows, 
\eqn\tensorpr{
\otimes_{\alpha=1}^L R_{\alpha}= \sum_s {\cal N}_{{\Lambda}_1, 
\cdots, \Lambda_L}^{\Lambda_s} R_s,}
where the integers ${\cal N}_{{\Lambda}_1, 
\cdots, \Lambda_L}^{\Lambda_s}$ can be easily computed by using repeatedly 
the Littlewood-Richardson rule. After taking into 
account the correction factors \rightfact, one finally obtains 
the following expression for the vev: 
\eqn\vevtl{
\langle \prod_{\alpha=1}^L W_{\Lambda_{\alpha}}^{(n,m)}\rangle 
=\sum_s {\cal N}_{{\Lambda}_1, 
\cdots, \Lambda_L}^{\Lambda_s} t^{{mn \over 2}(\sum_{\alpha=1}^L 
\kappa_{\Lambda_{\alpha}}-\kappa_{\Lambda_s})} 
\langle W_{\Lambda_s}^{(n,m)}\rangle,}
where the detailed expression for $\langle W_{\Lambda_s}^{(n,m)}\rangle$ 
was given in \lm, and $\kappa_{\Lambda}$ has been defined in \kap.
We will use \vevtl\ to give some explicit results on the link 
invariants in section 7. 
 
\subsec{Some predictions for the HOMFLY polynomial of links}

The result \frlinks\ gives a powerful structure theorem about the 
link invariants derived from Chern-Simons theory. Even in the case of 
links in the fundamental representation ({\it i.e.} the HOMFLY 
polynomial) one can obtain some highly non-trivial results. 
To extract the consequences 
of the above result, it is convenient to introduce some notation.  
Let's consider a link ${\cal L}$ of $L$ components ${\cal K}_{\alpha}$, 
$\alpha=1, \cdots, L$. The usual HOMFLY polynomial of this link 
will be denoted by $P_{\cal L}(q, \lambda)$, and it is related to the 
corrected Chern-Simons vev $\langle W({\cal L})\rangle$ as follows: 
\eqn\relatio{
\langle W({\cal L})\rangle=\lambda^{{\rm lk}({\cal L})}
\biggl( {\lambda^{1\over 2}-
\lambda^{-{1\over 2}}\over q^{1\over 2}-q^{-{1\over 2}}}\biggr) 
P_{\cal L}(q, \lambda).}
The structure theorem \frlinks\ says that
\eqn\linkst{
\langle W({\cal L})\rangle^{(c)}=(q^{-{1 \over 2}} -q^{{1\over 2}})^{L-2} 
\sum_Q \sum_{g\ge 0} {\widehat N}_{(\tableau{1}, \cdots, \tableau{1}),g,Q} 
\lambda^Q (q^{-{1\over 2}}-q^{{1\over 2}})^{2g}.}
We will first consider the simple case of a link of two components.
Using \twoc, we find that the HOMFLY polynomial of the link has the 
following structure: 
\eqn\twocstr{
P_{\cal L}(t, \lambda)=\sum_{g\ge 0} p^{\cal L}_{2g-1} 
(\lambda)(q^{1\over 2}-q^{-{1\over 2}})^{2g-1},} 
{\it i.e.} the lowest power of $q^{{1\over 2}}-q^{-{1\over 2}}$ is $-1$, 
and the powers are congruent to $-1$ mod 2. Moreover, if we denote the 
HOMFLY polynomial of the component knots by 
\eqn\compknot{
P_{{\cal K}_{\alpha}}(\lambda,q)=\sum_{g\ge 0}p^{{\cal K}_\alpha}_{2g}
(\lambda)
(q^{{1\over 2}}-q^{-{1\over 2}})^{2g},}
for $\alpha=1,2$, we find
\eqn\lmthtwo{
p_{-1}^{\cal L}(\lambda)=\lambda^{-{\rm lk}({\cal L})}(\lambda^{1\over2}-
\lambda^{-{1\over2}})p_0^{{\cal K}_1}(\lambda)p_0^{{\cal K}_2}(\lambda).} 
The last equation comes from the requirement that there are no powers of 
$(q^{{1\over 2}}-q^{-{1\over 2}})^{-2}$ in  
$\langle W({\cal L})\rangle^{(c)}$. The results \twocstr\ and \lmthtwo\ 
capture completely the algebraic structure of the HOMFLY polynomial of a 
two-component link, and reproduce the results of Lickorish 
and Millett \lickm.

We can generalize the above results for links with an arbitrary number 
of components $L$. 
We introduce again some notation. ${\cal L}_{\alpha_1, \cdots, \alpha_s}$ 
will denote the sublink of $s$ components obtained from the link ${\cal
L}$ 
by ``forgetting'' $L-s$ components, and $\{\alpha_1, 
\cdots, \alpha_s\} \subset \{1, \cdots, L\}$. The connected vev 
$\langle W({\cal L})\rangle^{(c)}$ is then given by 
the original vev together with some corrections involving 
products of vevs for sublinks: 
\eqn\conn{
\langle W({\cal L})\rangle^{(c)}=\langle W({\cal L})\rangle
-\sum_{{\alpha_L}=1}^L
\langle W({\cal L}_{\alpha_1, \cdots, \alpha_{L-1}})\rangle 
\langle W({\cal L}_{\alpha_L})\rangle 
+ \cdots.} By induction on the number of components, and using 
\conn\linkst, it is 
very easy to prove that the HOMFLY polynomial of the link has 
the following structure:
\eqn\link{
P_{\cal L}(q,\lambda)=
\sum_{g\ge0} p^{\cal L}_{2g+1-L} 
(\lambda) (q^{{1\over 2}}-q^{-{1\over 2}})^{2g+1-L},}
{\it i.e.} the lowest power of $q^{{1\over 2}}-q^{-{1\over 2}}$ is $1-L$. 
This has been also proved in \lickm. 
Due to our correction 
factor \modhom, it is convenient to 
introduce the following polynomials in $\lambda$:
\eqn\newpol{
{\widetilde p}^{{\cal L}_{\alpha_1, \cdots, \alpha_s}}_k 
(\lambda)={\lambda}^{{\rm lk}({\cal L}_{\alpha_1, \cdots, \alpha_s})} 
p^{{\cal L}_{\alpha_1, \cdots, \alpha_s}}_k 
(\lambda).} 
Finally, we will write 
\eqn\connagain{
\langle W({\cal L})\rangle^{(c)}=\biggl( {\lambda^{1\over2}-
\lambda^{-{1\over2}}\over q^{1\over2}-q^{-{1\over2}}}\biggr)\sum_{g\ge0} 
{\widetilde p}^{(c),{\cal L}}_{2g+1-L} 
(\lambda)(q^{1\over2}-q^{-{1\over2}})^{2g+1-L}.}  
The structure theorem \stf\ then states that
\eqn\strhom{
{\widetilde p}_{1-L}^{(c),{\cal L}}(\lambda)=
{\widetilde p}_{3-L}^{(c),{\cal L}}(\lambda)=
\cdots ={\widetilde p}_{L-3}^{(c),{\cal L}}(\lambda)=0.} This implies that 
the polynomials $p^{\cal L}_k(\lambda)$ of the HOMFLY polynomial of a
link, 
for $k=1-L, 3-L, \cdots, L-3$, are completely determined by the HOMFLY 
polynomial of its sublinks. As a first consequence, we 
find that ${\widetilde p}_{1-L}^{(c),{\cal L}}(\lambda)=0$ gives the 
generalization of \lmthtwo\ to an arbitrary link:
\eqn\lickmth{
p_{1-L}^{\cal L}(\lambda)=\lambda^{-{\rm lk}({\cal L})}(\lambda^{1\over2}- 
\lambda^{-{1\over2}})^{L-1} \prod_{\alpha=1}^L p_0^{{\cal
K}_{\alpha}}(\lambda).} 
This is easy to prove by induction on the number of components: 
since ${\widetilde p}_{1-L}^{(c), {\cal L}}(\lambda)=0$, one can extract
the 
coefficient of the lowest power of $q^{1\over2}-q^{-{1\over2}}$ in 
$\langle W({\cal L})\rangle$ from the terms involving products of vevs of 
sublinks in the expansion of the connected piece. One sees immediately
that 
the relevant part of these vevs is again the coefficient of 
the lowest power of $q^{1\over2}-q^{-{1\over2}}$. But because of the 
induction hypothesis, these in turn can be evaluated by factorization 
into their knots. This means that the coefficient 
${\widetilde p}_{1-L}^{\cal L}$  can be evaluated from $\prod_{\alpha=1}^L 
\langle W({\cal K}_{\alpha})\rangle$, and this proves \lickmth. 
This formula was in fact obtained by Lickorish and Millett in 
\lickm\ using the skein relation. In our context, this is 
just the simplest prediction of \strhom, which gives 
much more relations. For example, for links with $L=3$, 
the equality ${\widetilde p}_{0}^{(c), {\cal L}}(\lambda)=0$ implies that
\eqn\lasteq{
\eqalign{
 {\widetilde p}^{\cal L}_{0}(\lambda)&=(\lambda^{1\over2}-
\lambda^{-{1\over2}})(p_0^{{\cal K}_1}(\lambda) \,
{\widetilde p}_1^{{\cal L}_{23}}(\lambda) + {\rm perms}) \cr 
&-2(\lambda^{1\over2}-\lambda^{-{1\over 2}})^2(p_2^{{\cal K}_1}(\lambda)
\, p_0^{{\cal K}_2}(\lambda)\,p_0^{{\cal K}_3}(\lambda)+ {\rm
perms}).\cr}}
For links with more components, one obtains more complicated 
equations which can be summarized as in \strhom\ and 
give new results on the algebraic 
structure of the HOMFLY polynomial of links. 

\newsec{Lagrangian submanifold for torus links} 

As noted before, in order to describe the large $N$ closed
string dual for knot and link invariants, we need to construct
a suitable Lagrangian submanifold on the blown up conifold geometry
which corresponds to the D-branes after the conifold transition.
This has been done for the case of the trivial knot in
\ov . In this section we wish to generalize this construction
for certain links known as algebraic links (see 
for example \milnor\ for a discussion of
algebraic links). These are links that can be obtained by intersecting 
a plane curve in ${\bf C}^2$ 
\eqn\alg{
F(\zeta_1, \zeta_2)=0}
with a three-sphere $|\zeta_1|^2+ |\zeta_2|^2=R$. Torus links are in fact 
algebraic, since a torus link of type $(n,m)$ is described by the 
equation
\eqn\linkeq{
\zeta_1^n + \zeta_2^m =0.}
This can be easily seen by writing $\zeta_1 =r_1 {\rm e}^{i \theta_1}$, 
$\zeta_2 =r_2 {\rm e}^{i \theta_2}$. The number of components of this 
link is precisely $L={\rm gcd}(n,m)$. It is possible to obtain a more 
general algebraic equation describing the same torus link 
by taking the right hand side of \linkeq\ to be a polynomial in 
$\zeta_1,\zeta_2$ of lower degree (provided the radius $R$ of the 
three-sphere is big enough). It is important to notice 
that not all links and knots are algebraic. 

The equation \alg\ describes an algebraic curve in ${\bf C}^2$.
However on the blown up conifold geometry we are interested
in a 3 dimensional Lagrangian submanifold.  How does one
get a canonical such manifold from the above construction? 
There are two things that need to be changed here: first, we need
a Lagrangian submanifold rather than an algebraic one.  Secondly
we need a 3 dimensional manifold and not a 2-dimensional one.  

To remedy the first difficulty, notice that
one can 
perform a hyperK\"ahler rotation and obtain a two-dimensional (real) 
submanifold which is Lagrangian for the canonical K\"ahler form 
$\omega=(i/2)\sum_{k=1}^2 d\zeta_k \wedge d {\overline \zeta}_k$. 
The resulting equation is simply,
\eqn\lagr{
F({\rm e}^{i\theta/2}\zeta_1-{\rm e}^{-i\theta/2}{\overline \zeta}_2,
 {\rm e}^{i\theta/2}\zeta_2 + {\rm e}^{-i\theta/2}{\overline \zeta}_1)=0,}
whee $\theta$ is a real parameter.
To prove that this submanifold is Lagrangian, 
one considers the equation $dF=0$ and its complex conjugate, which 
read:
\eqn\df{
\eqalign{
& (\partial_1 F) ({\rm e}^{i\theta/2}d\zeta_1-
{\rm e}^{-i\theta/2}d{\overline \zeta}_2)=-(\partial_2 F)
({\rm e}^{i\theta/2}d\zeta_2 + {\rm e}^{-i\theta/2}d{\overline
\zeta}_1),\cr
& ( {\overline {\partial_1 F}}) ({\rm e}^{-i\theta/2}d{\overline \zeta}_1-
{\rm e}^{i\theta/2}d\zeta_2)=-({\overline {\partial_2 F}})
({\rm e}^{-i\theta/2}d{\overline \zeta}_2 + {\rm e}^{i\theta/2}d
\zeta_1).\cr}}
Assuming that $\partial_1 F \not=0$, and wedging the 
first equation with ${\rm e}^{i\theta/2}d\zeta_2 + {\rm
e}^{-i\theta/2}d{\overline \zeta}_1$ and the second equation with 
${\rm e}^{-i\theta/2}d{\overline \zeta}_2 + {\rm e}^{i\theta/2}d
\zeta_1$, we easily obtain that $\omega|_{F=0}=0$, for any $\theta$. 

To remedy the second difficulty we recall that
we are considering a Lagrangian submanifold in 
${\cal O}(-1)+{\cal O}(-1) \rightarrow {\bf P}^1$.   For each
point of ${\bf P}^1$ we have a copy of ${\bf C}^2$ and the ${\bf S}^3$
should be identified with the large ${\bf S}^3$ sitting in ${\bf C}^2$.
The
condition that the Lagrangian submanifold after transition corresponds
to a specific link before transition is simply that the Lagrangian
submanifold intersects the ${\bf S}^3$ at infinity along the corresponding
link.  This is precisely the case for the above construction.
Thus the above two dimensional construction can be viewed in principle
as a 2 dimensional slice of a 3d Lagrangian submanifold, where one
dimension of the Lagrangian submanifold projects to a curve in ${\bf
P}^1$.

In order to complete the story, we need to construct that third direction
of the Lagrangian submanifold.  The idea is to use the $SO(2)$ rotation
symmetry of ${\cal O}(-1)+{\cal O}(-1) \rightarrow {\bf P}^1$
(where one thinks of the fiber directions as two copies of spinors
and $SO(2)$ acts on ${\bf P}^1$ by the standard rotation),
and rotate the 2-dimensional Lagrangian submanifold to get a 3-dimensional
Lagrangian submanifold.  In particular consider the 3-dimensional
manifold which is given by the above Lagrangian submanifold, where
one interprets $\theta$ as the angle along the equator of ${\bf P}^1$.
In other words, the projection of the 3-manifold over ${\bf P}^1$
is given by the equator parameterized by $\theta$ and over
each point the fiber is the 2 dimensional Lagrangian submanifold
constructed above.
Note that the $\theta $ dependence in the above fibration
is consistent with the $SO(2)$ action on the fibers (being
spinors over the sphere).  This manifold makes sense if we get
the same fiber over $\theta =0$ and $\theta =2\pi$, which means that
this construction only makes sense for $F(\zeta_1,\zeta_2)=\pm
F(-\zeta_1,-\zeta_2)$.
We will now show that this 3 dimensional submanifold
is indeed Lagrangian, at least for a specific choice of metric on 
${\cal O}(-1)+{\cal O}(-1) \rightarrow {\bf P}^1$ which
is symplectically equivalent to that of the Ricci-flat metric.

 The 
Ricci-flat metric on this Calabi-Yau has a 
K\"ahler form that can be written as 
\eqn\kahl{
\omega={i \over 2}\partial {\overline \partial}K(U) + \omega_{{\bf P}^1}.}
In this equation, $\omega_{{\bf P}^1}$ is the usual symmetric
K\"ahler form on the 
sphere and $K(U)$ is a K\"ahler potential depending on the 
variable
\eqn\kalpot{
U=(1 + |z|^2)(|\zeta_1|^2 + |\zeta_2|^2),}
where $z$ is a complex coordinate on ${\bf P}^1$. 
Note that $U$ is simply the norm of the spinor bundles with respect
to the constant curvature metric on ${\bf P}^1$. 
Since we are only interested in the topological
string amplitudes only the K\"ahler class of the metric
should be relevant and not whether it is Ricci-flat or not.
For this any $K(U)$ will be sufficient.
We  take $K(U)=U$ (which
is not a Ricci-flat metric). In this case the K\"ahler form 
is given by
\eqn\kalfib{
{ i \over 2}\sum_{k=1}^2 \Bigl( (1+|z|^2)(d\zeta_k \wedge 
d{\overline \zeta}_k) + 
 {\overline z}\zeta_k dz\wedge d{\overline 
\zeta}_k +  z {\overline \zeta}_k d\zeta_k \wedge d{\overline 
z}\Bigr)+{\cal O}(dz\wedge d{\overline z}).}
where the terms involving $dz\wedge d{\overline z}$ are
clearly vanishing over the 3 dimensional manifold we have constructed.
Using this explicit expression for the K\"ahler form, it is 
easy to prove that for $z={\rm e}^{i \theta}$ ({\it i.e.} the 
equator of the sphere) the equation \lagr\ describes a 
Lagrangian submanifold which has the topology of a surface 
bundle over the equator. We have then obtained a Lagrangian 
submanifold in the resolved conifold corresponding to the 
algebraic link described by \alg. 

Notice that as explained above the equation \alg\ has monodromy as 
we go around the equator, therefore it is not always well-defined
for arbitrary torus links. For
 \linkeq, we must have $(n,m)$ both even or both odd. If this is 
the case, it is natural to propose that 
\lagr\ in fact describes the Lagrangian 
submanifold corresponding to the algebraic link \alg\ after the 
transition.     
As an evidence for the above proposal, note that this construction
for the case of 
$(n,m)=(1,1)$ which corresponds to the unknot reproduces
the result of \ov .  Also for the case of $(n,m)=(2,2)$, 
 which
corresponds to the Hopf link (and similarly for $(n,m)=(k,k)$),
the above construction agrees with a simple generalization
of the construction in \ov\ to this case. This provides further
evidence for our construction.

\newsec{Explicit results and examples} 

In this section we illustrate the results of the previous 
sections  
with explicit computations in Chern-Simons. The equations 
\fr\ and \frlinks\ give important and highly nontrivial 
structural predictions for 
the link invariants of Chern-Simons theory. The fact that 
they turn out to be true in all the cases that we have 
checked gives strong evidence for the 
D-brane degeneracy picture advocated in this paper. The results 
presented below are just an illustrative sample of all the 
examples that we have computed, involving many different 
torus knots and links. Since the resulting expressions 
are typically complicated, we have presented 
examples involving representations with a small 
number of boxes and 
knots and links with only a few crossings. Further examples 
are easily obtained with our results for torus links.

In the first subsection we present a computation of the new integer
invariants 
${\widehat N}_{R,g,Q}$ in the case of knots. The computation 
of the generating functions $f_R$ has been done in \lm\ for torus 
knots and in \rama\ for some other knots with up to nine crossings. 
{}From these generating functions, and using \relafs\ and the 
explicit expression of the matrix $M_{R\,R'}(q)$, we can 
extract the new integer 
invariants introduced in section 3. The fact that the functions 
${\widehat f}_R$ turn out to have the simple structure 
predicted in \tildefr\ is highly nontrivial from the 
point of view of Chern-Simons, since the entries of the 
inverse matrix of $M_{R\,R'}(q)$ are rational functions 
with very complicated denominators. The integrality 
property of the 
invariants ${\widehat N}_{R,g,Q}$ is clearly stronger than 
the integrality property of $N_{R,Q,s}$, and gives a 
powerful check of the arguments presented in section 3       

After presenting the results for knots, we consider the case 
of links and we  
illustrate the arguments presented in  
section 4. For links, the structural results that 
we have obtained are already 
very strong when all the components are in the 
fundamental representation, as we have seen in subsection 4.6. 
Our results also confirm that the proposal \gencor\ is 
the right prescription to obtain results compatible 
with the dual description in terms of topological strings. 
 
In principle, the integer invariants ${\widehat N}_{R,g,Q}$ can 
be computed on the gravity side, after the transition, by 
looking at the moduli space of Riemann surfaces with boundaries on the 
Lagrangian submanifold. It would be very interesting to compute some of
these 
numbers by using the explicit construction for torus links presented 
in section 5.

\subsec{Knots} 

We will first give some examples of the integer invariants 
${\widehat N}_{R,g,Q}$ for the trefoil knot, for representations of 
up to $3$ boxes. The values 
of the invariants can be read from equations (4.7), (4.13) and (4.21) of 
\lm, and are presented in tables 1-6. 

\smallskip
 
{\vbox{\ninepoint{
$$
\vbox{\offinterlineskip\tabskip=0pt
\halign{\strut
\vrule#
&
&\hfil ~$#$
&\hfil ~$#$
&\hfil ~$#$
&\hfil ~$#$
&\vrule#\cr
\noalign{\hrule}
&g
&Q=1/2
&3/2
&5/2
&
\cr
\noalign{\hrule}
&0
&2
&-3
&1
&\cr
&1
&1
&-1
&0
&\cr
\noalign{\hrule}}\hrule}$$}
\vskip - 7 mm
\centerline{{\bf Table 1:} The integers ${\widehat N}_{\tableau{1},g,Q}$
for 
the trefoil knot.}
\vskip7pt}
\noindent

\smallskip

{\vbox{\ninepoint{
$$
\vbox{\offinterlineskip\tabskip=0pt
\halign{\strut
\vrule#
&
&\hfil ~$#$
&\hfil ~$#$
&\hfil ~$#$
&\hfil ~$#$
&\hfil ~$#$
&\hfil ~$#$
&\vrule#\cr
\noalign{\hrule}
&g
&Q=1
&2
&3
&4
&5
&
\cr
\noalign{\hrule}
&0
&2
&-8
&12
&-8
&2
&\cr
&1
&1
&-6
&10
&-6
&1
&\cr
&2
&0
&-1
&2
&-1
&0
&\cr
\noalign{\hrule}}\hrule}$$}
\vskip - 7 mm
\centerline{{\bf Table 2:} The integers ${\widehat N}_{\tableau{2},g,Q}$
for 
the trefoil knot.}
\vskip7pt}
\noindent
\smallskip

{\vbox{\ninepoint{
$$
\vbox{\offinterlineskip\tabskip=0pt
\halign{\strut
\vrule#
&
&\hfil ~$#$
&\hfil ~$#$
&\hfil ~$#$
&\hfil ~$#$
&\hfil ~$#$
&\hfil ~$#$
&\vrule#\cr
\noalign{\hrule}
&g
&Q=1
&2
&3
&4
&5
&
\cr
\noalign{\hrule}
&0
&4
&-16
&24
&-16
&4
&\cr
&1
&4
&-20
&32
&-20
&4
&\cr
&2
&1
&-8
&14
&-8
&1
&\cr
&3
&0
&-1
&2
&-1
&0
&\cr
\noalign{\hrule}}\hrule}$$}
\vskip - 7 mm
\centerline{{\bf Table 3:} The integers ${\widehat N}_{\tableau{1 1},g,Q}$ 
for 
the trefoil knot.}
\vskip7pt}
\noindent

\medskip 

{\vbox{\ninepoint{
$$
\vbox{\offinterlineskip\tabskip=0pt
\halign{\strut
\vrule#
&
&\hfil ~$#$
&\hfil ~$#$
&\hfil ~$#$
&\hfil ~$#$
&\hfil ~$#$
&\hfil ~$#$
&\hfil ~$#$
&\hfil ~$#$
&\vrule#\cr
\noalign{\hrule}
&g
&Q=3/2
&5/2
&7/2
&9/2
&11/2
&13/2
&15/2
&
\cr
\noalign{\hrule}
&0
&2
&-18
&64
&-116
&114
&-58
&12
&\cr
&1
&1
&-21
&106
&-232
&251
&-131
&26
&\cr
&2
&0
&-8
&67
&-187
&227
&-121
&22
&\cr
&3
&0
&-1
&19
&-74
&103
&-55
&8
&\cr
&4
&0
&0
&2
&-14
&23
&-12
&1
&\cr
&5
&0
&0
&0
&-1
&2
&-1
&0
&\cr
\noalign{\hrule}}\hrule}$$}
\vskip - 7 mm
\centerline{{\bf Table 4:} The integers ${\widehat N}_{\tableau{3},g,Q}$ 
for 
the trefoil knot.}
\vskip7pt}
\noindent

\medskip
 
{\vbox{\ninepoint{
$$
\vbox{\offinterlineskip\tabskip=0pt
\halign{\strut
\vrule#
&
&\hfil ~$#$
&\hfil ~$#$
&\hfil ~$#$
&\hfil ~$#$
&\hfil ~$#$
&\hfil ~$#$
&\hfil ~$#$
&\hfil ~$#$
&\vrule#\cr
\noalign{\hrule}
&g
&Q=3/2
&5/2
&7/2
&9/2
&11/2
&13/2
&15/2
&
\cr
\noalign{\hrule}
&0
&11
&-99
&332
&-558
&507
&-239
&46
&\cr
&1
&15
&-201
&842
&-1627
&1612
&-796
&155
&\cr
&2
&7
&-164
&910
&-2080
&2275
&-1172
&224
&\cr
&3
&1
&-66
&528
&-1475
&1792
&-947
&167
&\cr
&4
&0
&-13
&171
&-614
&833
&-443
&66
&\cr
&5
&0
&-1
&29
&-148
&226
&-119
&13
&\cr
&6
&0
&0
&2
&-19
&33
&-17
&1
&\cr
&7
&0
&0
&0
&-1
&2
&-1
&0
&\cr
\noalign{\hrule}}\hrule}$$}
\vskip - 7 mm
\centerline{{\bf Table 5:} The integers ${\widehat N}_{\tableau{2 1},g,Q}$ 
for 
the trefoil knot.}
\vskip7pt}
\noindent
\medskip

{\vbox{\ninepoint{
$$
\vbox{\offinterlineskip\tabskip=0pt
\halign{\strut
\vrule#
&
&\hfil ~$#$
&\hfil ~$#$
&\hfil ~$#$
&\hfil ~$#$
&\hfil ~$#$
&\hfil ~$#$
&\hfil ~$#$
&\hfil ~$#$
&\vrule#\cr
\noalign{\hrule}
&g
&Q=3/2
&5/2
&7/2
&9/2
&11/2
&13/2
&15/2
&
\cr
\noalign{\hrule}
&0
&12
&-108
&352
&-568
&492
&-220
&40
&\cr
&1
&26
&-306
&1180
&-2136
&2006
&-950
&180
&\cr
&2
&22
&-366
&1740
&-3618
&3728
&-1864
&358
&\cr
&3
&8
&-230
&1431
&-3504
&3978
&-2066
&383
&\cr
&4
&1
&-79
&698
&-2077
&2603
&-1378
&232
&\cr
&5
&0
&-14
&200
&-761
&1057
&-561
&79
&\cr
&6
&0
&-1
&31
&-167
&259
&-136
&14
&\cr
&7
&0
&0
&2
&-20
&35
&-18
&1
&\cr 
&8
&0
&0
&0
&-1
&2
&-1
&0
&\cr
\noalign{\hrule}}\hrule}$$}
\vskip - 7 mm
\centerline{{\bf Table 6:} The integers ${\widehat N}_{\tableau{1 1
1},g,Q}$ 
for 
the trefoil knot.}
\vskip7pt}
\noindent

\medskip

In addition, we present also the the values of the integers ${\widehat
N}_{\tableau{1},g,Q}$, ${\widehat N}_{\tableau{2},g,Q}$ and ${\widehat
N}_{\tableau{1
1},g,Q}$ for the knots $4_1$, $5_1$ and $6_1$ shown in figure 1. 
The results are collected in tables 7-15. Knots $4_1$,
and $6_1$ are not torus knots. The values of the integer invariants have
been
obtained from the expressions for $f_{\tableau{1}}$, $f_{\tableau{2}}$ and
$f_{\tableau{1 1}}$ presented in \rama. The knot $5_1$ is a torus knot and
the
integer invariants 
have been obtained after using the general formula for this type
of knots provided in \lm. The following tables contain 
the values that take all these
integer invariants. Notice that for the amphicheiral knot $4_1$ the
results are
consistent with the fact that for this type of knots $f_R (q, \lambda)$ 
are invariant under
$q\rightarrow q^{-1}$, $\lambda\rightarrow \lambda^{-1}$. Using
 \lrschur\ and \fr\ it is easy to prove that amphicheiral knots 
satisfy
\eqn\amphi{
{\widehat N}_{R,g,Q}=(-1)^{\ell} {\widehat N}_{R^t,g,-Q},}
as one can see for $R=\tableau{1}$, $\tableau{2}$ and $\tableau{1 1}$ 
in tables 7-9.
  
\medskip
{\vbox{\ninepoint{
$$
\vbox{\offinterlineskip\tabskip=0pt
\halign{\strut
\vrule#
&
&\hfil ~$#$
&\hfil ~$#$
&\hfil ~$#$
&\hfil ~$#$
&\hfil ~$#$
&\vrule#\cr
\noalign{\hrule}
&g
&Q=-3/2
&-1/2
&1/2
&3/2
&
\cr
\noalign{\hrule}
&0
&1
&-2
&2
&-1
&\cr
&0
&0
&-1
&1
&0
&\cr
\noalign{\hrule}}\hrule}$$}
\vskip - 7 mm
\centerline{{\bf Table 7:} The integers ${\widehat N}_{\tableau{1},g,Q}$
for 
the figure-eight knot $4_1$.}
\vskip7pt}
\noindent

\smallskip

{\vbox{\ninepoint{
$$
\vbox{\offinterlineskip\tabskip=0pt
\halign{\strut
\vrule#
&
&\hfil ~$#$
&\hfil ~$#$
&\hfil ~$#$
&\hfil ~$#$
&\hfil ~$#$
&\hfil ~$#$
&\hfil ~$#$
&\hfil ~$#$
&\vrule#\cr
\noalign{\hrule}
&g
&Q=-3
&-2
&-1
&0
&1
&2
&3
&
\cr
\noalign{\hrule}
&0
&2
&-7
&9
&-6
&4
&-3
&1
&\cr
&1
&1
&-6
&9
&-5
&2
&-1
&0
&\cr
&2
&0
&-1
&2
&-1
&0
&0
&0
&\cr
\noalign{\hrule}}\hrule}$$}
\vskip - 7 mm
\centerline{{\bf Table 8:} The integers ${\widehat N}_{\tableau{2},g,Q}$
for 
the figure-eight knot $4_1$.}
\vskip7pt}
\noindent
\smallskip

{\vbox{\ninepoint{
$$
\vbox{\offinterlineskip\tabskip=0pt
\halign{\strut
\vrule#
&
&\hfil ~$#$
&\hfil ~$#$
&\hfil ~$#$
&\hfil ~$#$
&\hfil ~$#$
&\hfil ~$#$
&\hfil ~$#$
&\hfil ~$#$
&\vrule#\cr
\noalign{\hrule}
&g
&Q=-3
&-2
&-1
&0
&1
&2
&3
&
\cr
\noalign{\hrule}
&0
&1
&-3
&4
&-6
&9
&-7
&2
&\cr
&1
&0
&-1
&2
&-5
&9
&-6
&1
&\cr
&2
&0
&0
&0
&-1
&2
&-1
&0
&\cr
\noalign{\hrule}}\hrule}$$}
\vskip - 7 mm
\centerline{{\bf Table 9:} The integers ${\widehat N}_{\tableau{1 1},g,Q}$
for 
the figure-eight knot $4_1$.}
\vskip7pt}
\noindent
\smallskip

{\vbox{\ninepoint{
$$
\vbox{\offinterlineskip\tabskip=0pt
\halign{\strut
\vrule#
&
&\hfil ~$#$
&\hfil ~$#$
&\hfil ~$#$
&\hfil ~$#$
&\vrule#\cr
\noalign{\hrule}
&g
&Q=3/2
&5/2
&7/2
&
\cr
\noalign{\hrule}
&0
&3
&-5
&2
&
\cr
&1
&4
&-5
&1
&
\cr
&2
&1
&-1
&0
&
\cr
\noalign{\hrule}}\hrule}$$}
\vskip - 7 mm
\centerline{{\bf Table 10:} The integers ${\widehat N}_{\tableau{1},g,Q}$
for 
the  knot $5_1$.}
\vskip7pt}
\noindent
\smallskip

{\vbox{\ninepoint{
$$
\vbox{\offinterlineskip\tabskip=0pt
\halign{\strut
\vrule#
&
&\hfil ~$#$
&\hfil ~$#$
&\hfil ~$#$
&\hfil ~$#$
&\hfil ~$#$
&\hfil ~$#$
&\vrule#\cr
\noalign{\hrule}
&g
&Q=3
&4
&5
&6
&7
&
\cr
\noalign{\hrule}
&0
&20
&-80
&120
&-80
&20
&
\cr
&1
&60
&-260
&400
&-260
&60
&
\cr
&2
&69
&-336
&534
&-336
&69
&
\cr
&3
&38
&-221
&366
&-221
&38
&
\cr
&4
&10
&-78
&136
&-78
&10
&
\cr
&5
&1
&-14
&26
&-14
&1
&
\cr
&6
&0
&-1
&2
&-1
&0
&
\cr
\noalign{\hrule}}\hrule}$$}
\vskip - 7 mm
\centerline{{\bf Table 11:} The integers ${\widehat N}_{\tableau{2},g,Q}$ 
for  the  knot $5_1$.}
\vskip7pt}
\noindent
\smallskip

{\vbox{\ninepoint{
$$
\vbox{\offinterlineskip\tabskip=0pt
\halign{\strut
\vrule#
&
&\hfil ~$#$
&\hfil ~$#$
&\hfil ~$#$
&\hfil ~$#$
&\hfil ~$#$
&\hfil ~$#$
&\vrule#\cr
\noalign{\hrule}
&g
&Q=3
&4
&5
&6
&7
&
\cr
\noalign{\hrule}
&0
&30
&-120
&180
&-120
&30
&
\cr
&1
&115
&-490
&750
&-490
&115
&
\cr
&2
&176
&-819
&1286
&-819
&176
&
\cr
&3
&137
&-724
&1174
&-724
&137
&
\cr
&4
&57
&-365
&616
&-365
&57
&
\cr
&5
&12
&-105
&186
&-105
&12
&
\cr
&6
&1
&-16
&30
&-16
&1
&
\cr
&7
&0
&-1
&2
&-1
&0
&
\cr
\noalign{\hrule}}\hrule}$$}
\vskip - 7 mm
\centerline{{\bf Table 12:} The integers ${\widehat N}_{\tableau{1
1},g,Q}$ 
for  the  knot $5_1$.}
\vskip7pt}
\noindent

{\vbox{\ninepoint{
$$
\vbox{\offinterlineskip\tabskip=0pt
\halign{\strut
\vrule#
&
&\hfil ~$#$
&\hfil ~$#$
&\hfil ~$#$
&\hfil ~$#$
&\hfil ~$#$
&\hfil ~$#$
&\vrule#\cr
\noalign{\hrule}
&g
&Q=-3/2
&-1/2
&1/2
&3/2
&5/2
&
\cr
\noalign{\hrule}
&0
&1
&-1
&-1
&2
&-1
&
\cr
&1
&0
&-1
&0
&1
&0
&
\cr
\noalign{\hrule}}\hrule}$$}
\vskip - 7 mm
\centerline{{\bf Table 13:} The integers ${\widehat N}_{\tableau{1},g,Q}$
for 
the  knot $6_1$.}
\vskip7pt}
\noindent
\smallskip

{\vbox{\ninepoint{
$$
\vbox{\offinterlineskip\tabskip=0pt
\halign{\strut
\vrule#
&
&\hfil ~$#$
&\hfil ~$#$
&\hfil ~$#$
&\hfil ~$#$
&\hfil ~$#$
&\hfil ~$#$
&\hfil ~$#$
&\hfil ~$#$
&\hfil ~$#$
&\hfil ~$#$
&\vrule#\cr
\noalign{\hrule}
&g
&Q=-3
&-2
&-1
&0
&1
&2
&3
&4
&5
&
\cr
\noalign{\hrule}
&0
&2
&-4
&-1
&7
&-4
&-6
&15
&-13
&4
&
\cr
&1
&1
&-5
&2
&6
&-4
&-10
&25
&-19
&4
&
\cr
&2
&0
&-1
&1
&1
&-1
&-6
&13
&-8
&1
&
\cr
&3
&0
&0
&0
&0
&0
&-1
&2
&-1
&0
&
\cr
\noalign{\hrule}}\hrule}$$}
\vskip - 7 mm
\centerline{{\bf Table 14:} The integers ${\widehat N}_{\tableau{2},g,Q}$ 
for  the  knot $6_1$.}
\vskip7pt}
\noindent
\smallskip

{\vbox{\ninepoint{
$$
\vbox{\offinterlineskip\tabskip=0pt
\halign{\strut
\vrule#
&
&\hfil ~$#$
&\hfil ~$#$
&\hfil ~$#$
&\hfil ~$#$
&\hfil ~$#$
&\hfil ~$#$
&\hfil ~$#$
&\hfil ~$#$
&\hfil ~$#$
&\hfil ~$#$
&\vrule#\cr
\noalign{\hrule}
&g
&Q=-3
&-2
&-1
&0
&1
&2
&3
&4
&5
&
\cr
\noalign{\hrule}
&0
&1
&-2
&0
&3
&-4
&-12
&26
&-21
&6
&
\cr
&1
&0
&-1
&1
&1
&0
&-26
&60
&-46
&11
&
\cr
&2
&0
&0
&0
&0
&0
&-22
&50
&-34
&6
&
\cr
&3
&0
&0
&0
&0
&0
&-8
&17
&-10
&1
&
\cr
&4
&0
&0
&0
&0
&0
&-1
&2
&-1
&0
&
\cr
\noalign{\hrule}}\hrule}$$}
\vskip - 7 mm
\centerline{{\bf Table 15:} The integers ${\widehat N}_{\tableau{1
1},g,Q}$ 
for  the  knot $6_1$.}
\vskip7pt}
\noindent
\smallskip

\eject

\epsfverbosetrue

\ifig\laf{Some of the
knots and links considered in the paper. The trefoil knot $3_1$ and the
knot
$5_1$ are the torus knots $(3,2)$ and $(5,2)$,  in the notation used in
subsection 4.5. The figure-eight knot $4_1$ and the knot $6_1$ are not
torus
knots. The first one is amphicheiral. The two-component links $2_1^2$ and
$4_1^2$ correspond to the torus links $(2,2)$ and $(2,4)$ in the notation
used
in subsection 4.5. } {\epsfxsize 10.truecm \epsfysize
16.truecm\epsfbox{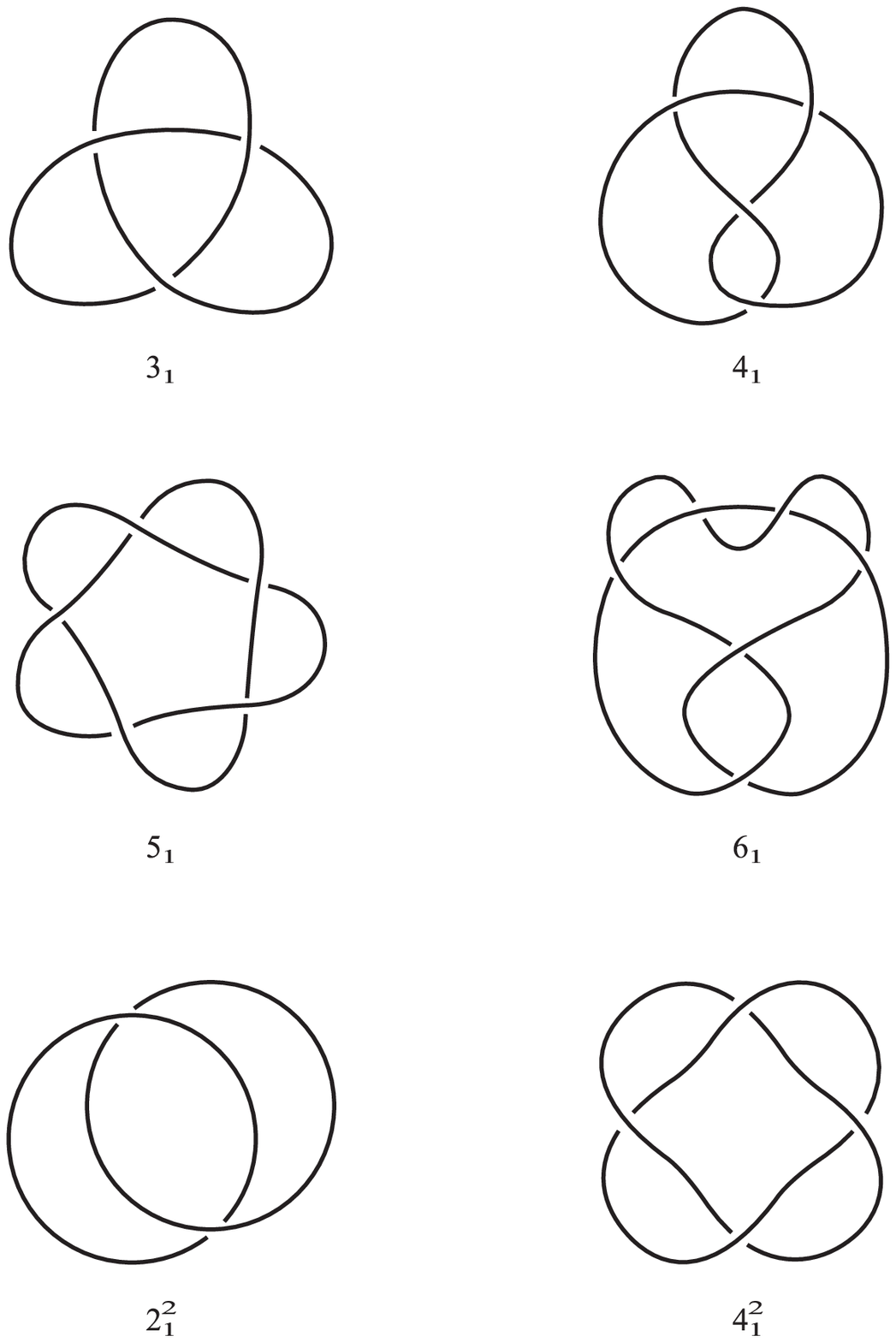}}

\subsec{Links}

We now give explicit results for some torus links.  
The simplest link is the Hopf link, which is the $(2,2)$ torus link. It
has 
two components which are both the unknot. Before presenting the results, 
it is worthwhile to show that the correction factor in \modhom\ is 
crucial to agree with the string predictions. The unnormalized HOMFLY 
polynomial of the Hopf link is given by: 
\eqn\homhopf{
\biggl( {\lambda^{1\over2}-\lambda^{-{1\over 2}} \over q^{1\over 2}
-q^{-{1\over 2}}}\biggr)
 \biggl( 
  {\lambda^{3\over 2}-\lambda^{1\over2} \over q^{1\over 2} -q^{-{1\over
2}}} 
-\lambda^{1\over2}(q^{1\over 2} -q^{-{1\over 2}})\biggr).}
The linking number of the two unknots is $-1$, and according to 
\modhom\ the vev $\langle {\rm Tr}\, U_1 \, {\rm Tr}\, U_2 \rangle$ 
should be equal to the unnormalized HOMFLY polynomial, times 
$\lambda^{-1}$. Without this factor, the connected vev 
\twoc\ does not have the structure predicted by \polyn. Once the right 
prescription has been used, one obtains:
\eqn\firstf{
f_{(\tableau{1},\tableau{1})}(q, \lambda)=-\lambda^{-1}(\lambda -1).}
Using all our previous results, we can compute the $f$'s up to 
order $\ell_1+\ell_2=4$. At order three, we find: 
\eqn\efesthree{
\eqalign{
f_{(\tableau{2}, \tableau{1})}(q, \lambda)=& -\lambda^{-1}(\lambda 
^{1\over 2}-\lambda^{-{1\over2}})q^{-{1\over 2}},\cr
f_{(\tableau{1 1}, \tableau{1})}(q, \lambda)=& \lambda^{-1}(\lambda 
^{1\over 2}-\lambda^{-{1\over2}})q^{1\over 2},\cr}}
while at order four we get:
\eqn\efesfour{
\eqalign{
f_{(\tableau{3}, \tableau{1})}(q, \lambda)=& -\lambda^{-2}(\lambda 
-1)q^{-1},\cr
f_{(\tableau{2 1}, \tableau{1})}(q, \lambda)=& \lambda^{-2}(\lambda 
-1),\cr
f_{(\tableau{1 1 1}, \tableau{1})}(q, \lambda)=& -\lambda^{-2} (\lambda 
-1)q,\cr
f_{(\tableau{2}, \tableau{2})}(q, \lambda)=& \lambda^{-2}(\lambda 
-1)q^{-2}(\lambda q +q^2 -q-1),\cr
f_{(\tableau{2}, \tableau{1 1})}(q, \lambda)=& -\lambda^{-2}(\lambda 
-1)^2,\cr
f_{(\tableau{1 1}, \tableau{1 1})}(q, \lambda)=& \lambda^{-2}(\lambda 
-1)(\lambda q -q^2 -q+1).\cr
}}
The next link in complexity is the torus link $(2,4)$, again 
a two-component link made of two unknots, with linking number $-2$. Up to 
order three, the results are:
\eqn\twofour{ 
\eqalign{
f_{(\tableau{1}, \tableau{1})}(q, \lambda)=& \lambda^{-1}q^{-1}(\lambda 
-1)(\lambda q-q^2-q-1),\cr
f_{(\tableau{2}, \tableau{1})}(q, \lambda)=& \lambda^{-{3\over 2}}q^{-{
5\over 2}}(\lambda 
-1)(1+q)(\lambda q-q^2 -1),\cr
f_{(\tableau{1 1}, \tableau{1})}(q, \lambda)=&-
\lambda^{-{3\over2}}q^{-{1\over 2}}(\lambda 
-1)(1+q)(\lambda q-q^2 -1).\cr
}}
Finally, we give a simple example of a link of 
three components: the  
link $(3,3)$, made up of three unknots. One finds:
\eqn\fthree{
f_{(\tableau{1}, \tableau{1},\tableau{1})}(q,\lambda)=
(q^{-{1\over 2}}- 
q^{1 \over 2})\Bigl\{ \lambda^{-{3\over 2}} 
(4 + (q^{-{1\over 2}} -q^{1\over 2})^2 )-\lambda^{ -{ 
1\over 2}} (5 +(q^{-{1\over 2}} -q^{1\over 2})^2) \Bigr\} ,}
in agreement with the structure results \stf\linkst.  

Using now the results of section 4, we can give the 
integer invariants ${\widehat N}_{(R_1,R_2),g,Q}$ for the Hopf 
link, for $\ell_1+\ell_2\le 4$. For $(R_1,R_2)=
(\tableau{1}, \tableau{1})$, we simply find
\eqn\hopfff{
{\widehat N}_{(\tableau{1}, \tableau{1}),0,0}=
-{\widehat N}_{(\tableau{1}, \tableau{1}),0,-1}=-1,}
and the rest of them are zero. For $(R_1,R_2)=
(\tableau{2}, \tableau{1})$, we find:
\eqn\hopfsf{
{\widehat N}_{(\tableau{2}, \tableau{1}),0,-1/2}=
-{\widehat N}_{(\tableau{2}, \tableau{1}),0,-3/2}=-1,}
and $N_{(\tableau{1 1}, \tableau{1}),g,Q}=0$ for all $g$, $Q$. 
For $(R_1,R_2)=
(\tableau{3}, \tableau{1})$, we have
 \eqn\hopftf{
{\widehat N}_{(\tableau{3}, \tableau{1}),0,-1}=
-{\widehat N}_{(\tableau{3}, \tableau{1}),0,-2}=-1,}
and $N_{(\tableau{2 1}, \tableau{1}),g,Q}=
N_{(\tableau{1 1 1}, \tableau{1}),g,Q}=0$ for all $g$, $Q$. 
For $(R_1,R_2)=
(\tableau{1 1}, \tableau{2})$, 
one simply has
\eqn\hopfantsym{
{\widehat N}_{(\tableau{1 1}, \tableau{2}),0,0}=-1,}
and $N_{(\tableau{1 1}, \tableau{1 1}),g,Q}=0$ for all $g$, $Q$. 
Finally, for $(R_1,R_2)=
(\tableau{2}, \tableau{2})$
the integer invariants are given in table 7.

\smallskip
 
{\vbox{\ninepoint{
$$
\vbox{\offinterlineskip\tabskip=0pt
\halign{\strut
\vrule#
&
&\hfil ~$#$
&\hfil ~$#$
&\hfil ~$#$
&\vrule#\cr
\noalign{\hrule}
&g
&Q=0
&1
&
\cr
\noalign{\hrule}
&0
&-3
&1
&\cr
&1
&-1
&0
&\cr
\noalign{\hrule}}\hrule}$$}
\vskip - 7 mm
\centerline{{\bf Table 16:} The integers 
${\widehat N}_{(\tableau{2},\tableau{2}),g,Q}$ for 
the Hopf link.}
\vskip7pt}
\noindent

For the torus link $(2,4)$, the integer invariants up to order 
four are presented in tables 17-24.
  
\smallskip
 
{\vbox{\ninepoint{
$$
\vbox{\offinterlineskip\tabskip=0pt
\halign{\strut
\vrule#
&
&\hfil ~$#$
&\hfil ~$#$
&\hfil ~$#$ 
&\hfil ~$#$
&\vrule#\cr
\noalign{\hrule}
&g
&Q=-1
&0
&1
&
\cr
\noalign{\hrule}
&0
&3
&-4
&1
&\cr
&1
&1
&-1
&0
&\cr
\noalign{\hrule}}\hrule}$$}
\vskip - 7 mm
\centerline{{\bf Table 17:} The integers 
${\widehat N}_{(\tableau{1},\tableau{1}),g,Q}$ for 
the link $(2,4)$. }
\vskip7pt}
\noindent

\smallskip

{\vbox{\ninepoint{
$$
\vbox{\offinterlineskip\tabskip=0pt
\halign{\strut
\vrule#
&
&\hfil ~$#$
&\hfil ~$#$
&\hfil ~$#$ 
&\hfil ~$#$
&\vrule#\cr
\noalign{\hrule}
&g
&Q=-3/2
&-1/2
&1/2
&
\cr
\noalign{\hrule}
&0
&6
&-9
&3
&\cr
&1
&5
&-6
&1
&\cr 
&2
&1
&-1
&0
&\cr
\noalign{\hrule}}\hrule}$$}
\vskip - 7 mm
\centerline{{\bf Table 18:} The integers 
${\widehat N}_{(\tableau{2},\tableau{1}),g,Q}$ for 
the link $(2,4)$. }
\vskip7pt}}
\noindent

\smallskip
 
{\vbox{\ninepoint{
$$
\vbox{\offinterlineskip\tabskip=0pt
\halign{\strut
\vrule#
&
&\hfil ~$#$
&\hfil ~$#$
&\hfil ~$#$ 
&\hfil ~$#$
&\vrule#\cr
\noalign{\hrule}
&g
&Q=-3/2
&-1/2
&1/2
&
\cr
\noalign{\hrule}
&0
&2
&-3
&1
&\cr
&1
&1
&-1
&0
&\cr 
\noalign{\hrule}}\hrule}$$}
\vskip - 7 mm
\centerline{{\bf Table 19:} The integers 
${\widehat N}_{(\tableau{1 1},\tableau{1}),g,Q}$ for 
the torus link $(2,4)$. }
\vskip7pt}

\smallskip
{\vbox{\ninepoint{
$$
\vbox{\offinterlineskip\tabskip=0pt
\halign{\strut
\vrule#
&
&\hfil ~$#$
&\hfil ~$#$
&\hfil ~$#$ 
&\hfil ~$#$
&\vrule#\cr
\noalign{\hrule}
&g
&Q=-2
&-1
&0
&
\cr
\noalign{\hrule}
&0
&10
&-16
&6
&\cr
&1
&15
&-20
&5
&\cr 
&2
&7
&-8
&1
&\cr
&3
&1
&-1
&0
&\cr
\noalign{\hrule}}\hrule}$$}
\vskip - 7 mm
\centerline{{\bf Table 20:} The integers 
${\widehat N}_{(\tableau{3},\tableau{1}),g,Q}$ for 
the torus link $(2,4)$. }
\vskip7pt}
\noindent

 \smallskip
 
{\vbox{\ninepoint{
$$
\vbox{\offinterlineskip\tabskip=0pt
\halign{\strut
\vrule#
&
&\hfil ~$#$
&\hfil ~$#$
&\hfil ~$#$ 
&\hfil ~$#$
&\vrule#\cr
\noalign{\hrule}
&g
&Q=-2
&-1
&0
&\cr
\noalign{\hrule}
&0
&5
&-8
&3
&\cr
&1
&5
&-6
&1
&\cr 
&2
&1
&-1
&0
&\cr
\noalign{\hrule}}\hrule}$$}
\vskip - 7 mm
\centerline{{\bf Table 21:} The integers 
${\widehat N}_{(\tableau{2 1},\tableau{1}),g,Q}$ for 
the torus link $(2,4)$. }
\vskip7pt}
\noindent

The integers ${\widehat N}_{(\tableau{1 1 1}, \tableau{1}),g,Q}$ all
vanish.

\smallskip

{\vbox{\ninepoint{
$$
\vbox{\offinterlineskip\tabskip=0pt
\halign{\strut
\vrule#
&
&\hfil ~$#$
&\hfil ~$#$
&\hfil ~$#$
&\hfil ~$#$
&\hfil ~$#$
&\vrule#\cr
\noalign{\hrule}
&g
&Q=-2
&-1
&0
&1
&
\cr
\noalign{\hrule}
&0
&48
&-93
&54
&-9
&\cr
&1
&106
&-172
&72
&-6
&\cr
&2
&99
&-137
&39
&-1
&\cr
&3
&47
&-57
&10
&0
&\cr
&4
&11
&-12
&1
&0
&\cr
&5
&1
&-1
&0
&0
&\cr
\noalign{\hrule}}\hrule}$$}
\vskip - 7 mm
\centerline{{\bf Table 22:} The integers ${\widehat N}_{(\tableau{2}, 
\tableau{2}),g,Q}$ 
for 
the torus link $(2,4)$ .}
\vskip7pt}
\noindent
 
\smallskip

{\vbox{\ninepoint{
$$
\vbox{\offinterlineskip\tabskip=0pt
\halign{\strut
\vrule#
&
&\hfil ~$#$
&\hfil ~$#$
&\hfil ~$#$
&\hfil ~$#$
&\hfil ~$#$
&\vrule#\cr
\noalign{\hrule}
&g
&Q=-2
&-1
&0
&1
&
\cr
\noalign{\hrule}
&0
&26
&-47
&24
&-3
&\cr
&1
&45
&-67
&23
&-1
&\cr
&2
&30
&-38
&8
&0
&\cr
&3
&9
&-10
&1
&0
&\cr
&4
&1
&-1
&0
&0
&\cr
\noalign{\hrule}}\hrule}$$}
\vskip - 7 mm
\centerline{{\bf Table 23:} The integers ${\widehat N}_{(\tableau{2}, 
\tableau{1 1}),g,Q}$ 
for 
the torus link $(2,4)$.}
\vskip7pt}
\noindent  \smallskip

{\vbox{\ninepoint{
$$
\vbox{\offinterlineskip\tabskip=0pt
\halign{\strut
\vrule#
&
&\hfil ~$#$
&\hfil ~$#$
&\hfil ~$#$
&\hfil ~$#$
&\hfil ~$#$
&\vrule#\cr
\noalign{\hrule}
&g
&Q=-2
&-1
&0
&1
&
\cr
\noalign{\hrule}
&0
&12
&-21
&10
&-1
&\cr
&1
&16
&-22
&6
&0
&\cr
&2
&7
&-8
&1
&0
&\cr
&3
&1
&-1
&0
&0
&\cr
\noalign{\hrule}}\hrule}$$}
\vskip - 7 mm
\centerline{{\bf Table 24:} The integers ${\widehat N}_{(\tableau{1 1}, 
\tableau{1 1}),g,Q}$ 
for 
the torus link $(2,4)$.}
\vskip7pt}
\noindent

\bigskip
\centerline{\bf Acknowledgments}
\bigskip

We would like to thank K. Hori, M. Liu, J. Maldacena, G. Moore,
H. Ooguri and C. Taubes for useful conversations. 
M.M. would like to thank the High 
Energy Theory Group at Harvard for hospitality. J.M.F.L. 
would like to thank the
Department of Physics at the University of Maryland, where part of
this work was done, for hospitality. The work of J.M.F.L. 
is supported in part by
DGICYT under grant PB96-0960. The work of M.M. is supported by DOE grant
DE-FG02-96ER40959. The work of C.V. is partially  supported by NSF grant
PHY-98-02709.

\listrefs
\bye